\documentclass[sigconf,authorversion,nonacm]{acmart}
\usepackage[utf8]{inputenc}
\usepackage[english]{babel}
\usepackage{amsthm}
\usepackage{graphicx}
\usepackage{hyperref}
\usepackage{listings}
\usepackage{amsmath}
\usepackage{multirow}

\definecolor{fraunhoferDarkGreen}{RGB}{12, 78, 63}
\definecolor{lightGreen}{RGB}{197, 246, 235}
\definecolor{darkBlue}{RGB}{0, 110, 146}
\definecolor{lightBlue}{RGB}{211, 241, 250}

\title{Tackling Consistency-related Design Challenges of Distributed Data-Intensive Systems -- An Action Research Study}\titlenote{This is a post-print of an article published in the conference proceedings of the 15th ACM/IEEE International Symposium on Empirical Software Engineering and Measurement (ESEM) (ESEM 21). It is available online at: \textbf{https://doi.org/10.1145/3475716.3475771}. \newline
\textbf{Please cite this version:} Susanne Braun, Stefan Deßloch, Eberhard Wolff, Frank Elberzhager, and Andreas Jedlitschka. 2021. Tackling Consistency-related Design Challenges of Distributed Data-Intensive Systems -- An Action Research Study. In ACM/IEEE International Symposium on Empirical Software Engineering and Measurement (ESEM) (ESEM ’21), October 11–15, 2021, Bari, Italy. ACM, NewYork, NY, USA, 12 pages. https://doi.org/10.1145/3475716.34757711}

\author{Susanne Braun}
\affiliation{
\institution{Fraunhofer IESE}
\city{Kaiserslautern} 
\country{Germany}}
\email{susanne.braun@iese.fraunhofer.de}

\author{Stefan Deßloch}
\affiliation{
\institution{TU Kaiserslautern}
\city{Kaiserslautern} 
\country{Germany}}
\email{stefan.dessloch@cs.uni-kl.de}

\author{Eberhard Wolff}
\affiliation{
\institution{INNOQ}
\country{Germany}}
\email{eberhard.wolff@innoq.com}

\author{Frank Elberzhager}
\affiliation{
\institution{Fraunhofer IESE}
\city{Kaiserslautern} 
\country{Germany}}
\email{frank.elberzhager@iese.fraunhofer.de}

\author{Andreas Jedlitschka}
\affiliation{
\institution{Fraunhofer IESE}
\city{Kaiserslautern} 
\country{Germany}}
\email{andreas.jedlitschka@iese.fraunhofer.de}

\copyrightyear{2021} 
\acmYear{2021} 
\setcopyright{acmlicensed}\acmConference[ESEM '21]{ACM / IEEE International Symposium on Empirical Software Engineering and Measurement (ESEM)}{October 11--15, 2021}{Bari, Italy}
\acmBooktitle{ACM / IEEE International Symposium on Empirical Software Engineering and Measurement (ESEM) (ESEM '21), October 11--15, 2021, Bari, Italy}
\acmPrice{15.00}
\acmDOI{10.1145/3475716.3475771}
\acmISBN{978-1-4503-8665-4/21/10}

\begin{document}



\begin{abstract}
\textbf{Background:} Distributed data-intensive systems are increasingly designed to be only eventually consistent. Persistent data is no longer processed with serialized and transactional access, exposing applications to a range of potential concurrency anomalies that need to be handled by the application itself. Controlling concurrent data access in monolithic systems is already challenging, but the problem is exacerbated in distributed systems. To make it worse, only little systematic engineering guidance is provided by the software architecture community regarding this issue. \textbf{Aims:} In this paper, we report on our study of the effectiveness and applicability of the novel design guidelines we are proposing in this regard. \textbf{Method:} We used action research and conducted it in the context of the software architecture design process of a multi-site platform development project. \textbf{Results:} Our hypotheses regarding effectiveness and applicability have been accepted in the context of the study. The initial design guidelines were refined throughout the study. Thus, we also contribute concrete guidelines for architecting distributed data-intensive systems with eventually consistent data. The guidelines are an advancement of Domain-Driven Design and provide additional patterns for the tactical design part. \textbf{Conclusions:}  Based on our results, we recommend using the guidelines to architect safe eventually consistent systems. Because of the relevance of distributed data-intensive systems, we will drive this research forward and evaluate it in further domains.
\end{abstract}

\maketitle

% \vspace*{-0.5\baselineskip}
\section{Introduction}
\label{sec:intro}
% 1 Page -> 1,5 Pages
With the success of cloud-native and mobile applications, modern software architectures have become highly distributed. Whether microservices \cite{wolff2016microservices} or mobile applications supporting disconnected operation -- apps and services increasingly need to be resilient to network partitions, high latency, or temporary unavailability of services. Therefore, more and more data is being replicated across different replication nodes (replicas).
% even across different geographical regions.
Updates to replicated data are propagated asynchronously, without global coordination or transactional guarantees, i.e., the ACID (\underline{A}tomicity, \underline{C}onsistency, \underline{I}solation, \underline{D}urability) \cite{DBLP:journals/csur/HaerderR83} guarantees. This can result in concurrency and consistency anomalies like lost updates \cite{DBLP:conf/sigmod/BerensonBGMOO95}, which need to be handled by the application itself. 

Consider a cloud-based digital workspace offering apps and cognitive services to remote teams. This digital workspace provides APIs designed to facilitate the development of third-party workspace apps.
% for serving of specific needs of different types of knowledge work.
Platform services and third-party app services need to be loosely coupled, autonomous, independent deployment units and resilient to failures of other services. Pivotal data such as tasks are stored by different apps and services of the ecosystem.
% so that these can be self-contained, autonomous and resilient to the unavailability of other apps and services. 
For example, teams will agree on a particular task management system as the primary tool for self-organization. In addition, a team dashboard app displays information relevant to the team, such as
% forecasts of the prospective availability of teammates, or 
a team mood barometer. The team dashboard also hosts a daily meeting buddy app presenting a summary of the tasks the user has worked on since the last daily meeting. The user can indicate when they start to work on a new task using the buddy app. Such updates need to be propagated to the primary task management tool.
% and result in the task being moved to the "in-progress" lane.
Similarly, updates to tasks in the primary tool need to be propagated to the buddy app. Consider a user who starts working on a task and shares this information via 
% the functionality of 
the buddy app. At about the same time, the product owner uses the primary tool to move this task to the blocked lane and also adds a comment providing the rationale. As the task management system and the buddy app do not have a shared infrastructure for storage, there is no central instance for coordinating these concurrent updates. Instead, the result will be the existence of two conflicting versions of the task. The implementation of a generic merge function for conflicting task versions is more challenging than it might look at first sight. For example, we can build the super set of all comments from both versions in order to reconcile the changes to the list of task comments \cite{DBLP:journals/cacm/Vogels09}. This strategy is fine most of the time. However, it can lead to the effect that deleted comments re-appear \cite{DBLP:journals/cacm/Vogels09}. More challenging is the question of what the final task status is supposed to be: blocked or in progress? We could also apply syntactic conflict resolution strategies such as the last-writer-wins rule to solve this problem, but then one update will be lost, which is probably worse. 

In our experience, it is nearly impossible to implement correct merge logic in practice. 
% Thus, in case of conflicts correctness is usually traded off for other quality attributes, in particular availability. 
Some conflicts can be correctly resolved in retrospect, but require additional information such as the common base version of the conflicting versions. It would be even more beneficial to know the original intents of the concurrent update operations causing the conflicting versions. Therefore, a potential solution can be to use an operation-based update propagation interface \cite{DBLP:series/synthesis/2008Terry} and, in addition, design updating operations to form a lattice so that conflicts can be excluded by design. Lattices are the foundation of Conflict-Free Replicated Data Types (CRDTs) \cite{DBLP:conf/sss/ShapiroPBZ11}, ACID 2.0 \cite{DBLP:conf/cidr/HellandC09}, and the CALM theorem \cite{DBLP:journals/cacm/HellersteinA20} (CALM stands for \underline{C}onsistency \underline{A}s \underline{L}ogical
\underline{M}onotonicity). But this research has not yet been successfully transferred into practice.

% Another example is support for disconnected operation. Members of remote teams are often mobile workers using mobile devices under conditions where connectivity is not guaranteed. To support disconnected operation data needs to be replicated to the mobile devices so that mobile workers can continue to update work artifacts during network outages. Updates are propagated as soon as bandwidth is sufficient. The longer the network outage the higher the chance for conflicting updates.

In our case studies \cite{DBLP:conf/wicsa/NaabBLHEMCK15, DBLP:journals/software/BraunCN16, DBLP:conf/icsa/BraunD20}, we learned that practitioners often underestimate the complexity of developing custom update propagation and conflict resolution schemes. They merely address consistency and concurrency challenges with trial-and-error. Another problem is that only few experts are able to address these challenges in a targeted way. A summary of our assumptions based on anecdotal observations is given in \cite{DBLP:conf/eurosys/BraunBE21}. We also learned that consistency- and concurrency-related challenges need to be considered already during the functional decomposition of the overall application domain into different sub-domains and, in particular, during the design of the sub-domain models. Alas, only little  engineering guidance is available that is aimed at the design of domain models that are safe under concurrent execution of updates running in parallel on multiple nodes of a distributed system. We refer to domain models consisting (partly or fully) of eventually consistent data that can be updated safely and concurrently at different replication nodes as ``safe eventually consistent domain models''. Considering our empirical observations and the results of related studies \cite{DBLP:conf/icsa/HasselbringS17, DBLP:journals/ese/ScavuzzoNA18, godefroid2008concurrency, DBLP:conf/esem/AhmedB18, DBLP:journals/cloudcomp/PardonPZ18}, we are convinced that concrete design guidelines in this regard would be beneficial.

We conducted an action research study  \cite{DBLP:books/sp/Staron20} in a realistic research setting, focusing on the following research question: 

\begin{itemize}
\item
  \textbf{RQ: How can we design safe eventually consistent domain models in a targeted way?} 
\end{itemize}

We provide the following contributions:

\begin{itemize}
\item
  An initial empirical validation of novel design guidelines we are proposing in this regard.
\item
  Comprehensive design guidelines extending the tactical design part of Domain-Driven Design (DDD) \cite{DBLP:books/EvansDDD}. These guidelines specifically address con\-sis\-ten\-cy- and concurrency-related design challenges of distributed data-intensive systems.
% \item
%   A theory explaining how practitioners usually handle con\-sis\-ten\-cy- and concurrency-related design challenges of distributed data-intensive systems.
\end{itemize}

The design guidelines initially emerged as lessons learned from two other case studies \cite{DBLP:conf/wicsa/NaabBLHEMCK15, DBLP:conf/icsa/BraunD20} and were extensively revised in the course of the study. We applied action research in the context of the software architecture design
process of a medium-sized platform development project with an overall development budget of 220 person-months. The responsibility of the project team is to deliver a digital workspace platform as described above. 
% The overall application domain of the remote work platform is decomposed into
% different apps and services with corresponding sub-domain models such as task management, a digital assistant for remote teams, and so on. 

The remainder of this paper is structured as follows: We provide an overview of related work in Section  \ref{sec:relatedWork}. 
% We present our theory explaining how practitioners usually handle consistency- and concurrency-related design challenges in Section \ref{sec:theory}. 
Our design method is outlined in Section \ref{sec:designMethod}, including a description of our design guidelines. Section \ref{sec:researchDesign} details our research design and Section \ref{sec:results} presents the results and discusses them.  
% We discuss our work in Section \ref{sec:discussion}, provide further learnings in Section \ref{sec:learnings} and 
Validity threats are examined in Section \ref{sec:validity}. We summarize key results and conclude the paper in Section \ref{sec:summmary-conclusions}. 

% \vspace*{-\baselineskip}
\section{Background and Related Work}
\label{sec:relatedWork}
% 1 Page
Kleppmann defines an application as data-intensive ``if data is its primary challenge -- the quantity of data, the complexity of data, or the speed at which it is changing -- as opposed to compute-intensive, where CPU cycles are the bottleneck'' \cite{DBLP:books/oreilly/Kleppmann2014}. Other works often focus on the first and the third challenge, while the increasing complexity of data is often overlooked. Increased complexity of data comes with increased complexity of our domain models and 
% correspondingly increased complexity of the software and systems 
the software we need to architect and build. This is an area where DDD shines. Therefore, we will first give a short introduction to DDD as relevant background and then discuss related work. 

% \subsection{Domain-Driven Design}
% \label{sec:DDD}
% Domain-Driven Design is all about the achievement of one primary goal in
% software engineering: reducing complexity in software. 
% DDD in particular addresses the complexity that is already inherent in the domain or the business of the user itself \cite{DBLP:books/EvansDDD}. 
DDD suggests that complex domain designs be 
based on a model -- the domain model \cite{DBLP:books/EvansDDD}. 
% Large software projects, involve different stakeholder groups, and these different groups
% speak different business languages and formulate different requirements
% for the domain model. 
Trying to maintain one global (canonical) model often fails in
practice as these models become too complex and ambiguous. 
% Ambiguity often arises when different stakeholder groups have the same terms in
% their business language, but use it with different meanings. 
Therefore,
one central concept of DDD are bounded contexts. A bounded
context defines a clear boundary within which one specific unified 
model is valid. It further defines which business language is valid
within this context -- the so-called ubiquitous language, used in models and code and spoken by stakeholders and developers.
% It is important that this clearly defined language
% is spoken by the domain experts as well as the software engineers, and
% that this language is also consistently used in the model and the code.
% Therefore, DDD refers to it as a ubiquitous language. 
% A bounded
% context is usually maintained by
% one team. 
In a microservice architecture, one bounded context usually
corresponds to one microservice. The strategic design of DDD helps to come up with the right bounded context cut. The tactical
design deals with the design of the domain model itself.

% \vspace*{-\baselineskip}
\subsection{Consistency and Concurrency}
\label{sec:background}
Today's data-intensive systems are highly distributed in order to achieve higher availability and scalability by exploting increased parallelism. This trend will continue as ``CPU clock speeds are barely increasing, but multi-core processors are standard, and networks are getting faster'' \cite{DBLP:books/oreilly/Kleppmann2014}. But increased ``distributed parallelism'' poses new challenges for software engineers. In this section, we provide an overview of the engineering support already available to date.

ACID transactions \cite{DBLP:journals/csur/HaerderR83} provide extensive guarantees. Proper use of transaction isolation levels \cite{DBLP:conf/sigmod/BerensonBGMOO95} can ensure that concurrent data access is equivalent to serial execution (serializability), completely eliminating the possibility of concurrency anomalies. Isolation and atomicity are the foundations of consistency in the original sense of ACID: If the transaction program itself is correct, data can always be in a consistent state, as related updates can be executed -- from the perspective of the application -- in isolation and in one instant \cite{DBLP:books/mk/GrayR93}. ACID consistency is concerned with the internal consistency of application data, specifically the validity of application invariants and constraints.

Distributed transactions \cite{DBLP:journals/vldb/BreitbartGS92} and sagas \cite{DBLP:conf/sigmod/Garcia-MolinaS87, RichardsonSagaPattern} span a number of sub-trans\-act\-ions executed in different systems, e.g., different services in a service-oriented architecture. Distributed transactions and sagas provide the ``all-or-nothing'' guarantee for related updates that need to be executed together 
% for maintaining  ACID consistency 
in a distributed system. Distributed transactions solve this with ACID atomicity, sagas with compensation. Note that they do not address concurrency issues arising from eventually consistent replication as outlined above.

As for database replication, it is well known ``that transactional replication is unstable as the workload scales up'' \cite{DBLP:conf/sigmod/GrayHOS96}. This limitation is still valid today, except for very few databases offering ACID transactions also on replicated data, e.g., Google Spanner \cite{corbett2013spanner, brewer2017spanner}. 
%However, if Spanner replicates data across different regions, remote replicas are read-only and serve potentially stale data \cite{spannerDocumentationReplication}. 
Brewer proposed BASE (\underline{B}asically \underline{A}vailaible, \underline{S}oft State, \underline{E}ventual Consistency) \cite{DBLP:conf/podc/Brewer00} 
% as opposite concept of ACID 
in this regard
% ACID and BASE are at opposite ends of the whole consistency-availability design spectrum
\cite{DBLP:conf/podc/Brewer00, brewer2012cap}. 
BASE trades off isolation for availability by favoring eventual consistency, meaning that updates are propagated only asynchronously. The period
between the successful execution of an update on one replica
and the full propagation to all other replicas
is called the inconsistency window \cite{DBLP:journals/cacm/Vogels09}. Eventual consistency only ``guarantees'' that all replicas will eventually converge to the same state if update activity ceases \cite{DBLP:series/synthesis/2008Terry}. Unlike eventual consistency, strong consistency \cite{DBLP:series/synthesis/2008Terry} guarantees that the last update is always returned to any subsequent access. 
% Eventual consistency is one of a multitude of variants of weak consistency \cite{TODO}. The opposite concept of weak consistency is strong consistency \cite{TODO}, guaranteeing that the last update is always returned to any subsequent access. ACID consistency includes strong consistency \cite{brewer2012cap}. 
% BASE also forfeits ACID durability as during inconsistency windows concurrent operations might unknowingly violate invariants. These operations need to be rolled back or compensated during reconciliation and are therefore not durable. 
During inconsistency windows, applications might read stale data 
% and perform further actions
% based on the outdated values they have read 
or even update the same data
concurrently, resulting in concurrency anomalies and conflicts that need to be handled by the application itself. Implementation of custom code for conflict resolution is a huge source of human error \cite{DBLP:conf/icsa/Braun17, DBLP:books/oreilly/Kleppmann2014}, which is why some researchers have proposed exploiting ``semantic tricks'' like commutativity of update operations to prevent conflicts from happening in the first place, e.g., \cite{DBLP:conf/sigmod/GrayHOS96}. Weikum et al. introduced Multilevel Transactions for increased concurrency in relational databases \cite{DBLP:books/mk/elmagarmid92/WeikumS92}. Helland proposed ACID 2.0 \cite{DBLP:conf/cidr/HellandC09} but does not show how to implement it in practice. Shapiro et al. did extensive research on Conflict-Free Replicated Datatypes (CRDTs) \cite{DBLP:conf/sss/ShapiroPBZ11}, but currently developers are limited to a small number of existing data types. Implementation of custom CRDTs
% , in particular, the CRDT merge function, 
is intellectually demanding and has not been transferred into practice. Notably, there is the CALM theorem \cite{DBLP:journals/cacm/HellersteinA20}, which defines a class of problems that can be safely implemented in a distributed system and do not require global coordination. 

To conclude, recent approaches focus on application-level consistency and compatibility of ``higher-level'' operations (e.g., domain operations) instead of conflict relations of lower-level read and write operations on the infrastructure layer. Also note how ACID consistency differs from strong consistency and eventual consistency: The latter are always subject to replicated data and provide different guarantees regarding the visibility of updates at distinct replicas. In contrast, ACID consistency 
% guarantees are much more powerful, 
includes strong consistency \cite{brewer2012cap} and all-time validity of domain invariants and constraints. 
% or a federation of (potentially heterogeneous) database management systems.

% \vspace*{-\baselineskip}
\subsection{Software Engineering Guidance}
\label{sec:seGuidanceConsistency}
Microservices blur the boundaries in the discussion of transactional replication vs. eventual consistency and distributed transactions vs. sagas because they usually need both replication and distribution of data. Regarding replication, pivotal domain objects such as an order object in an e-commerce system often have to be stored by multiple microservices. Even though these duplicated domain objects are modeled individually in different bounded contexts, they still represent the same real-world concept and are therefore kind of replicated at the application level. 
% Allowing different models in different bounded contexts -- which is for good reasons -- just adds complexity to update propagation as some additional mappings between different object representations have to be performed. Depending on the quality of the bounded context cut, the degree of replication is expected to vary. With an optimal cut driven by functionality, we assume that the degree of replicated domain objects can be minimized, but rarely completely eliminated. 
At the same time, distributed transactions or sagas are needed as chances are low that use cases will always require updates to be executed only in single microservices. 
% Use cases touching domain invariants that span multiple microservices, can be an indicator for a bad microservice cut. However, refactoring the microservice cut might not always be an option and therefore distributed transactions still should be considered and balanced with the complexity of implementing correct compensating sub-transactions of the alternative Saga-based approach. 
% Due to these special characteristics of microservices, the term eventual consistency is often used imprecisely and mixed up with ACID consistency (e.g. in \cite{FowlerMicroserviceTradeoffs, DBLP:conf/icsa/HasselbringS17, DBLP:journals/cloudcomp/PardonPZ18}). 
Correspondingly, most of the available software engineering guidance on ``data consistency'' is related to microservices, e.g., \cite{richardson_microservices_2018, FowlerMicroserviceTradeoffs, DBLP:conf/icsa/HasselbringS17, DBLP:journals/cloudcomp/PardonPZ18}.
% , providing concrete design patterns e.g. the saga pattern \cite{RichardsonSagaPattern}. 

Practitioners are discussing the use of event sourcing \cite{FowlerEventSourcing} for asynchronous propagation of updates between microservices \cite{RichardsonEventSourcingPattern}, similar to the idea of state-machine replication \cite{DBLP:journals/csur/Schneider90}. As event sourcing was originally not intended for this, it seems to be a violation of the information hiding principle, as the event log should be private for an event-sourced object. An alternative pattern is to combine CQRS (\underline{C}ommand \underline{Q}uery \underline{R}esponsibility \underline{S}egregation) \cite{cqrs} with the transactional outbox pattern for propagation of commands (updates) between microservices \cite{RichardsonOutboxPattern, RichardsonCQRSPattern}. In the context of CQRS, eventual consistency originally referred to the inconsistencies between the command model and the query model. In non-distributed settings, this is uncritical, as the processing of commands and asynchronous updating of the query model can be synchronized with local transactions. However, this is not the case in distributed systems with application-level data replication.
% across different systems. 
To summarize: though these patterns can be used for update propagation, they do not offer any  guidance for addressing  potentially emerging consistency and concurrency anomalies.
% Using it the way it was originally proposed \cite{cqrs}, asynchronous updating of the query model can be synchronized by local transactions.

Apart from microservice architectures, Kleppmann \cite{DBLP:books/oreilly/Kleppmann2014} can serve as a starting point, but does not provide in-depth software architecture guidance such as design patterns. To maintain ACID consistency, DDD has the aggregates pattern. Aggregates are clusters of domain objects that need to be updated atomically so that aggregate-internal consistency and domain invariants can be maintained. DDD therefore suggests updating aggregates in the context of ACID transactions \cite{DBLP:books/EvansDDD}. To avoid performance issues stemming from database isolation, DDD further suggests modeling aggregates to be only as large as strictly necessary and preferably updating only one aggregate in one transaction \cite{DBLP:books/EvansDDD, vernon2013implementingDDD}.

Unfortunately, our literature research did not yield empirical studies that directly investigated the efficacy of engineering methods for the safe architecting of distributed data-intensive systems with eventually consistent data. Other studies, already quoted in the introduction \cite{DBLP:conf/icsa/HasselbringS17, DBLP:journals/ese/ScavuzzoNA18, godefroid2008concurrency, DBLP:conf/esem/AhmedB18, DBLP:journals/cloudcomp/PardonPZ18}, are only distantly related but assert our assumptions: The resulting design challenges are demanding and more design guidance would be appreciated by practitioners. 
% Further, reproducibility of concurrency defects is a big problem in practice \cite{godefroid2008concurrency, DBLP:conf/esem/AhmedB18}, which is why our research aims at avoiding concurrency and consistency-related defects in the first place. 

% \vspace*{-0.5\baselineskip}
\section{Design Method}
\label{sec:designMethod}
% 1,5 Pages Gesamt
% Zunächst Einleitung und Überblick
% Grafik einfügen
% viertel Seite inkl. Grafik
Our design guidelines\footnote{The version of the guidelines used during the study is provided in the replication package: https://doi.org/10.6084/m9.figshare.14988405} aim at achieving two goals: 
\begin{enumerate}
\def\labelenumi{\arabic{enumi}.}
\item
  \textbf{Goal: semantic compatibility of domain operations} to allow them to run concurrently and conflict-free at different replicas.
\item
  \textbf{Goal: minimizing the chance of conflicts} with an optimized design of domain objects.
\end{enumerate}

Our design guidelines therefore consist of two documents, each dedicated to one goal:  the ``Domain Operations Design Guide'' assists in achieving the first goal, and the ``Domain Objects Design Guide'' aids in reaching the second goal. The final document structure of our design guides is given in Table \ref{tab:designGuidelines}. Each document provides some 20 pages of content.

\begin{table*}
    \caption{Structure of the Design Guidelines}
    \label{tab:designGuidelines}
    \begin{minipage}{\textwidth}
    \begin{center}
    \begin{tabular}{p{0.485\linewidth} | p{0.485\linewidth}}
    \toprule
        \textbf{Domain Operations Design Guide} & \textbf{Domain Objects Design Guide}\\
        \hline
      Introduction \& Prerequisites & Introduction \& Prerequisites\\
    %   \hline
      Best Practice -- Design for ``Tolerance to Partial Execution Order'' & Aggregate Taxonomy \\
      Domain Operations Compatibility Relation & 3 Trivial Aggregate Classes \\
      Best Practice -- Design for Incremental Updates &  $\rightarrow$ Classification Criteria, Examples, Guiding Questions \\
      Best Practice -- Design for True Blind Updates & 3 Non-Trivial Aggregate Classes \\
      3 Compatibility Anti-Patterns & $\rightarrow$ Classification Criteria, Examples, Guiding Questions \\
      Best Practice -- Design for Domain Invariant Consistency & 5 Data Model Design Best Practices \\
      Best Practice -- Consider Durability Requirements & Cheat Sheet \\
      Cheat Sheet & \\
    \bottomrule
    \end{tabular}
    \end{center}
    \end{minipage}
\end{table*}

% Alternative ways to design for semantic compatibility are illustrated in detail in our domain operations design guide which we'll outline in section \ref{sec:domainOperationsDesign}. The technical prerequisites enabling exploitation of these compatibility relations are presented in section \ref{sec:technicalPrerequisites}. 
% Next, our design guidelines target at reducing the chance for conflicts with an optimized design of domain objects. 
% Corresponding best practices and design patterns are given in the domain objects design guide which we'll summarize in section \ref{sec:domainObjectsDesign}. 
% An overview on the overall engineering approach is given in figure \ref{fig:engApproach}.
% \begin{figure}
%     \centering
%     \includegraphics[width=\columnwidth]{figures/Overview-Approach.pdf}
%     \caption{Overall Engineering Approach}
%     \label{fig:engApproach}
% \end{figure}

\subsection{Domain Operations Design Guide}
\label{sec:domainOperationsDesign}
% viertel Seite
% Semantic compatibility of domain operations is usually equated with commutativity. Examples often used are withdraw and deposit operations on bank accounts. Because of the commutativity property of integer addition, naive implementations of withdraw and deposit also commute. However, real-world domain logic is usually complex. Upon closer examination, it commonly turns out that domain operations are actually not commutative. For example, adding a single domain invariant to assure that the account balance remains above a certain threshold renders the withdraw operation non-commutative.
% % with itself and the deposit operation. 
% Thus, our study aimed at exploring alternative compatibility relations that go beyond commutativity. 

The domain operations design guide assists software engineers in designing compatible domain operations. Compatible operations can safely be re-executed on all nodes (potentially on a different state) and still
correctly produce the same intended updates as during the original execution. We refer to the latter property as the ability of domain operations to tolerate partial execution order. A detailed description of this property and how to achieve it during operational design is given in the guide. 
In addition, compatible design of operations guarantees the eventual absence of lost updates \cite{DBLP:conf/eurosys/BraunBE21}. In this regard, we distinguish two basic types of updates: incremental updates and true blind updates. 
% Domain operations perform incremental updates if, prior to updating the state, the current state has to be evaluated first in order to derive the updated state value. 
% Examples of incremental updates are counter increments or list appends. 
% Incremental updates are prone to lost updates.
%Obviously, incremental updates are prone to lost updates: If the "evaluation phase" is executed during an inconsistency window and the current state is updated concurrently at another replica, the concurrent update is lost for the moment. 
% However, if domain operations are designed according to our guidelines and an update propagation \& reconciliation scheme, as described in section \ref{sec:technicalPrerequisites}, is used, lost updates of incremental updates can be fixed in retrospect during reconciliation. 
% True blind updates are the opposite of incremental updates: The state is blindly overwritten and the updated state value is independent of the current state value. 
% If blind updates are true blind updates -- in the sense that the updated state value is, in fact, \textit{not} an increment of the current state value -- then lost updates can be excluded in the first place, based on the definition of the lost update phenomenon itself \cite{DBLP:conf/sigmod/BerensonBGMOO95}. 
The guide makes use of best practices using UML models and code snippets for illustrating the two types. It also uses anti-patterns to further facilitate learning of the design methodology (see Table \ref{tab:designGuidelines}).
% e.g. to clearly distinguish true blind updates from technical blind updates, which do lead to lost updates and are often found in anemic domain models \cite{FowlerAnemicDomainModel}. 

\subsection{Domain Objects Design Guide}
\label{sec:domainObjectsDesign}
Depending on the domain, it might not always be feasible to come up with a design where all domain operations are compatible with each other. In this case, an optimized design of the domain objects can reduce the chance of conflicts. An optimal aggregate cut can lead to a significant reduction of a model's conflict potential. Similar to coming up with an optimal bounded context cut, designing an optimal aggregate cut is demanding. If distinct domain objects have differing update characteristics, this can be an indicator that these domain objects need not be updated atomically and therefore also need not be clustered within the same aggregate boundary \cite{DBLP:books/EvansDDD, vernon2013implementingDDD}. But ``differing update characteristics'' is vague and does not provide concrete guidance on how to apply this principle systematically during the design process. To approach this in a more targeted way, we provide concrete classification criteria related to aggregate update behavior and use these to derive a taxonomy of six different aggregate classes \cite{DBLP:conf/icsa/BraunD20}. We mainly distinguish trivial and non-trivial aggregates. Trivial aggregates, such as immutable and derived aggregates, rule out the possibility of conflicts, whereas non-trivial aggregates have a certain potential for conflicts depending on their concrete class membership. For each aggregate class, we provide concrete classification criteria, examples, and further guiding questions to facilitate classification. Based on the taxonomy, we also provide concrete design best practices for the construction of aggregate cuts with minimum conflict potential (see Table \ref{tab:designGuidelines}). 

\section{Research Design}
\label{sec:researchDesign}
% 1,5 Pages
The aim of the study is to evaluate software architecture design guidelines supporting practitioners in tackling consistency-related design challenges of distributed data-intensive systems. Real-world software architectures are long-lived and complex. It is thus nearly impossible to measure any effects related to real-world software architecture problems in controlled experiments \cite{DBLP:journals/ese/FalessiBCK10}. In case studies, researchers are usually only allowed to observe and not permitted to drive forward any changes to established software architecture practices, e.g., by proposing a new design technique. We therefore decided to conduct our study under the action research paradigm \cite{DBLP:books/sp/Staron20}, as 
it allows researchers to collaborate with
practitioners in the context of real projects. 
% Researchers are allowed to intervene to support
% practitioners. Such interventions are called actions. Action research is
% always timeboxed and conducted iteratively in cycles \cite{DBLP:books/sp/Staron20}.
% Each cycle has the following phases: (1) diagnosing the problem, (2)
% planning actions, (3) taking actions, (4) evaluating actions, and (5)
% learning and theory building \cite{DBLP:books/sp/Staron20}.

% \vspace*{-\baselineskip}
\subsection{Context}
\label{sec:context}
The case company, Fraunhofer-Gesellschaft, is a large group of applied research institutions in Germany obliged to realize one third of its turnover from projects with industry customers. The studied organization, Fraunhofer IESE, is doing applied software engineering research. It also offers IT consultancy services to industry customers. It employs more than 100 computer science researchers and eleven full-time developers. The problem of missing design guidelines was originally identified at the case organization during a software development project conducted between 2014 and 2016 on behalf of a large manufacturer of agricultural machinery \cite{DBLP:conf/wicsa/NaabBLHEMCK15, DBLP:journals/software/BraunCN16}. 
%After this cooperation the first author continued to do research on this topic in the context of a PhD. 
%The initial version of the domain objects design guide is a distillation of the results that emerged from the subsequent analysis of  different domain models of three case studies \cite{DBLP:conf/icsa/BraunD20}. A literature survey and the study of related work such as current research on CRDTs and ACID 2.0 lead to the first version of the domain operations design guide. 

At the end of 2018, a multi-site platform development project was started at the case organization together with two other medium-sized software development companies. The goal of this project is to develop a minimum-viable product (MVP) of a digital workspace as outlined in the introduction (see Section \ref{sec:intro}). 
% offering intelligent and cognitive services to remote teams until mid of 2021. The APIs of this platform are designed to facilitate the development of third-party workspace apps and services% as an enabler for the formation of an ecosystem of integrated digital workspace apps and services from different providers.
% Platform services and third-party services need to be loosely coupled, independent deployment units and resilient to failures of other services.  
The workspace platform was designed according to the microservices architecture paradigm. The idea was to develop open host services and corresponding published languages \cite{DBLP:books/EvansDDD} for the platform-wide asynchronous propagation of updates to pivotal aggregates, such as tasks or meetings, which were expected to be independently stored and updated in different platform services or apps. 
As the application domain is the remote work domain, web clients should also be able to support disconnected operation. Thus, client-specific bounded context models needed to be replicated to edge devices with an eventually consistent replication scheme. Correspondingly, the design challenges addressed by our guidelines were a potential issue in this project and qualified the project as validation context. 

The overall software development project budget is 220 person-months; the budget of the development team at the case organization is 92 person-months. At the date of writing, the platform project has more than 30 git repositories and more than 20 microservices. Development activities started in 2019 with sprint zero and a small team consisting of a product owner, a lead software architect, and two full-stack developers. In 2020, the regular multi-site development process started applying a combination of Agile and Lean principles. At the case organization, three full-time-equivalent developers and one full-time-equivalent UX designer as well as one full-time-equivalent product owner were staffed to the project. Other experts, such as data scientists, as well as data privacy experts were temporarily involved depending on the project phase and the tasks. 
%During sprint zero a senior architect was responsible for elicitation of the primary architecture drivers and the design of the major architecture principles. Regarding data privacy and security issues further specialists were involved for the development of data privacy and security concepts. 
In Q1 and Q2 of 2021, the development team at the case organization was reduced to one part-time product owner, one part-time UX designer, and two part-time full-stack developers who spent approx. 60\%, 30\%, 50\%, and 80\%, respectively, of their working time in the development team. 
The two development teams at the industry sites each had three developers and one project manager working in the team on average. Due to delays, the second industry team's development activities could only begin in March 2021 and will continue after the first official release of the MVP in July 2021 until the end of 2021.
Multi-site development activities are coordinated in four-week sprints. The development team at the case organization organized its work internally in two-week sprints and aligned them with the four-week sprints of the multi-site development process.

 Due to the complexity of the platform development project itself, all requirements resulting in the need to handle eventual consistency were postponed in the course of sprint zero.
%  , such as support for disconnected operation of web clients. The idea to design open host services for asynchronous propagation of updates between microservices was also deferred. 
 For any data replicated at the application level between different microservices, users have to configure a leading app.
%  for management of pivotal domain objects such as tasks. 
 Other services and apps need to rely on the availability of the synchronous REST API of the leading app regarding the execution of update operations. 
 In Q4 of 2020, the disconnected operation requirement was prioritized again. Action research was used to develop eventually consistent bounded context models for individual workspace apps selected to support disconnected operation on edge devices.

% \vspace*{-0.5\baselineskip}
\subsection{Theoretical Framework}
\label{sec:theoFramework}
% Designing distributed data-intensive systems with eventual consistency is demanding as a lot of complexity is shifted from the infrastructure layer into the domain layer. Complexity that used to be hidden and handled by infrastructure now needs to be addressed in the domain and application layers. For example, it is necessary to explicitly consider during the design process whether it is safe to execute different domain operations concurrently at different replication nodes. In case of conflicts, appropriate conflict resolution strategies need to be designed. Whether conflicts, as well as concurrent execution of domain operations, result in severe anomalies, such as lost updates is highly dependent on the domain, its use cases and its semantics. Also, an optimal design according to the strategic and tactical design of DDD can contribute to reduce the need for synchronous global coordination, such as distributed transactions based on 2PC. Therefore, DDD is a natural fit and provides a good foundation for providing additional guidance specifically addressing challenges related to all kinds of eventually-consistent data replication. 
We designed our study based on the assumption that a practitioner's approach is usually characterized by  trial-and-error, drastic underestimation of the inherent complexity of custom synchronization schemes, and a lack of understanding of the required underlying concepts. A summary of our assumptions and observations gained in past case studies can be found in \cite{DBLP:conf/eurosys/BraunBE21}. We therefore hypothesize that our guidelines meet a real need of practitioners.

\subsection{Summary of Research Cycles}
\label{sec:sumCycles}
\begin{figure}
    \centering
    \includegraphics[width=\columnwidth]{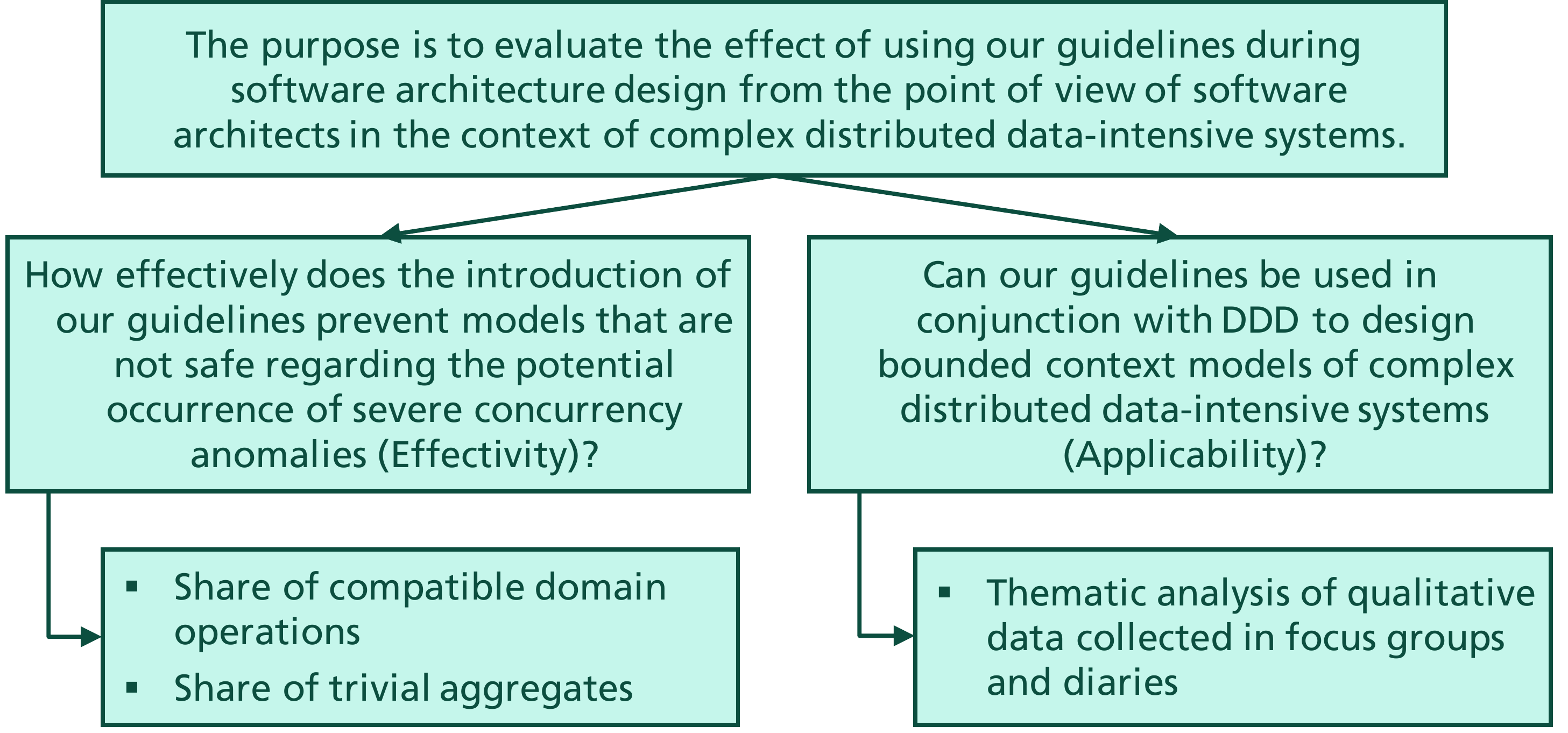}
    \caption{GQM research goals}
    \label{fig:gqm}
\end{figure}

In October 2020, the action team set off to identify best practices and design patterns for the design of safe eventually consistent domain models. Action research is
always timeboxed and conducted iteratively in cycles \cite{DBLP:books/sp/Staron20}.
% to support disconnected operation of selected workspace apps. 
Each action research cycle has the following phases: (1) diagnosing the problem, (2)
planning actions, (3) taking actions, (4) evaluating actions, and (5)
learning and theory building \cite{DBLP:books/sp/Staron20}.
We ran two cycles between late October 2020 and early March 2021. Each cycle had a duration of between six and eights weeks. The research question of both cycles was: 
\begin{itemize}
\item
  \textbf{RQ: How can we design safe eventually consistent domain models in a targeted way?} 
\end{itemize}
 The research goals, the corresponding sub-questions, as well as the metrics applied during the evaluation phases are shown in the GQM template \cite{caldiera1994goal} presented in Figure \ref{fig:gqm}. We focused on measuring the effectiveness and applicability of the guidelines; the corresponding hypotheses are given in Table \ref{tab:hypotheses}. The design of an eventually consistent domain model is safe if severe concurrency anomalies, such as lost updates, can be prevented by design. Thus, if all domain operations are compatible and can therefore run concurrently and conflict-free on different nodes, the model is safe by design. Correspondingly, one metric for measuring the safety of a domain model is its share of compatible domain operations. A second measure is the share of trivial aggregates of the model, as 
% the occurrence of 
conflicts can be excluded in connection with trivial aggregates. To assess applicability, we conducted a thematic analysis \cite{thematicAnalyisis} on qualitative data collected in focus groups and diaries.

\begin{table*}
    \caption{Research Hypotheses}
    \label{tab:hypotheses}
    \begin{minipage}{\textwidth}
    \begin{center}
    \begin{tabular}{p{0.06\linewidth} p{0.89\linewidth}}
    \toprule
        \textbf{H\textsubscript{Effect1}} & The introduction of our guidelines results in the  design of safe eventually consistent domain models, so that the occurrence of concurrency anomalies, in particular lost updates, is prevented by design.\\
        %  \hline
        \textbf{H\textsubscript{App1}} & Professional software engineers trained in DDD are able to apply our guidelines as part of the tactical design.\\
        % \hline
        \textbf{H\textsubscript{App2}} & Professional software engineers trained in DDD applying our guidelines as part of the tactical design are able to come up with a reasonable functional decomposition of the eventually consistent domain model.\\
        % \hline
        \textbf{H\textsubscript{App3}} & Additional complexity of eventually consistent domain models caused by being compliant with our guidelines is acceptable in practice (and adequate for addressing the inherently complex task of controlling concurrent data access on different replicas).\\
    \bottomrule
    \end{tabular}
    \end{center}
    \end{minipage}
\end{table*}

\begin{figure*}
    \centering
    \includegraphics[width=\textwidth]{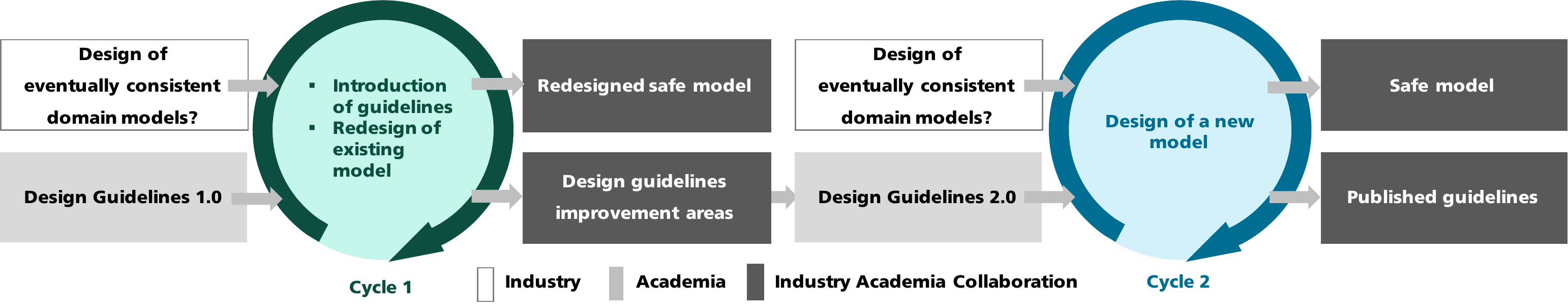}
    \caption{Overview of Research Cycles}
    \label{fig:overviewCycles}
\end{figure*}

Figure \ref{fig:overviewCycles} provides an overview of how the study was conducted and shows the inputs from academia and industry as well as the outputs of the industry-academia collaboration of each cycle. Table \ref{tab:overviewCycles} contains a summary of each cycle (the first row refers to the first cycle, the second row to the second cycle).

\begin{table*}
    \caption{Summary of Research Cycles - Cycle 1 in first row - Cycle 2 in second row}
    \label{tab:overviewCycles}
    \begin{minipage}{\textwidth}
    \begin{center}
    \begin{tabular}{p{0.18\linewidth} p{0.18\linewidth} p{0.18\linewidth} p{0.18\linewidth} p{0.18\linewidth}}
    \toprule
        \textbf{Diagnosing} & \textbf{Action Planning} & \textbf{Action Taking} & \textbf{Evaluation} & \textbf{Learning}\\
        \hline
        % \textcolor{fraunhoferGreen}{}
        \color{fraunhoferDarkGreen}
        Focus group 1 accepted our observations from past case studies \cite{DBLP:conf/eurosys/BraunBE21}. The participants would approach the design with trial-and-error. How can we instead design safe eventually consistent models in a targeted way? &
        \color{fraunhoferDarkGreen}
       The researchers suggested presenting the design guidelines to the action team. The action team agreed to use the guidelines to re-design the model of the team mood barometer app to support disconnected operation of the app. & 
        \color{fraunhoferDarkGreen}
        The design guidelines were presented to the action team during a 1-hour online meeting. The action team also got written guidelines in textbook style. Two developers independently created a re-design of the model. & 
         \color{fraunhoferDarkGreen}
        The share of compatible domain operations as well as the share of trivial aggregates was significantly higher in the re-designs than in the existing baseline bounded context model. & 
         \color{fraunhoferDarkGreen}
        The domain objects design guide required improvement in several areas (e.g., the classification criteria were partly ambiguous). The compatibility relation for domain operations was hard to understand and required a major revision.\\
        \hline
        \color{darkBlue}
        As the guidelines required improvement in several areas, the problem was the same as in the first cycle: How to design safe eventually consistent domain models in a targeted way? &
        \color{darkBlue}
        The researchers planned to provide the revised design guidelines to the action team. The action team planned to design a bounded context model from scratch using the revised guidelines. &
        \color{darkBlue}
        The action team designed a larger bounded context model for backlog management, backlog grooming, and asynchronous conduction of planning pokers using the revised written guidelines. &
        \color{darkBlue}
        In the resulting design, 21 out of 22 domain operations were compatible. The share of trivial aggregates was around 40\%. Due to the characteristics of domain data, this share could not have been increased meaningfully. &
        \color{darkBlue}
        The action team appreciated the value of the guidelines, which were now perceived to be easy to understand and apply. Due to their benefits, the guidelines should be published internally and externally.\\
    \bottomrule
    \end{tabular}
    \end{center}
    \end{minipage}
\end{table*}

\subsection{Data Collection and Analysis Methods}
\label{sec:DataCollAnalysis}
In this section, we describe how we collected and analyzed quantitative and qualitative data. First, we provide general information regarding the participants of our study. Next, we explain how data was collected and analyzed in the different phases of each cycle\footnote{We provide a replication package at: https://doi.org/10.6084/m9.figshare.14988405}.

Regarding the participants, action research distinguishes different actors: the members of the action
team and the members of the reference group \cite{DBLP:books/sp/Staron20}. The action team is responsible for
planning, executing, and evaluating the research \cite{DBLP:books/sp/Staron20}. The reference
group is not involved in action planning, action taking, and
evaluation, but has practical experience with the problem and can
therefore provide objective feedback to evaluation results and reduce
biases \cite{DBLP:books/sp/Staron20}. An overview of the actors of this study is given in Table \ref{tab:actors}. The action team was staffed with two researchers (R). The first author was the first researcher. She is the responsible product owner of the platform MVP and the designer of the action research study. She is also the author of the design guidelines. The fourth author was the second researcher. He had not worked in the previous case studies that led to the original design guidelines. His role was to adopt a neutral position. Correspondingly, he was the moderator of all focus groups and feedback sessions. He also reviewed any transcripts prior to their analysis by the first researcher. He regularly chatted informally with the other action team members as participants sometimes try to please researchers during focus groups. Both researchers have more than ten years of professional experience. The designer of the action research study had been employed in
industry for more than five years before joining the case organization, working as a software engineer and software architect. In addition, two full-stack developers (D) from the case organization were part of the action team. Both were members of the development team since sprint zero. The first full-stack developer has eight years of professional experience as a developer and also took over the role of the lead software architect of the project in mid-2020. The second developer has four years of professional experience as a developer. Two members of the leadership team (professional experience between 12 and 20 years) were in the reference group (L). Both reference group members had worked in at least one of the previous case studies mentioned above (thus had an understanding of the problem) and still regularly work in strategic projects with industry customers in the role of a software architect.

\begin{table*}
    \caption{Actors}
    \label{tab:actors}
    \begin{minipage}{\textwidth}
    \begin{center}
    \begin{tabular}{p{0.05\linewidth} p{0.18\linewidth} p{0.3\linewidth} p{0.18\linewidth} p{0.18\linewidth}}
    \toprule
        \textbf{Actor} & \textbf{Actor Type} & \textbf{Job Title} & \textbf{Work \mbox{Experience}} & \textbf{Project Role} \\
        \hline
        R1 & Action Team Member & Expert Data-Intensive Systems Design & 12 years & Product Owner\\
        R2 & Action Team Member & Expert Scientist Quality Assurance & 16 years & - \\
        D1 & Action Team Member & Senior Developer & 8 years & Lead Software Architect\\
        D2 & Action Team Member & Senior Developer & 4 years & Lead DevOps Engineer\\
        L1 & Reference Group & Department Head & 12 years & - \\
        L2 & Reference Group & Division Manager & 20 years & -\\
    \bottomrule
    \end{tabular}
    \end{center}
    \end{minipage}
\end{table*}

% \paragraph{Diagnosing Phase}
\subsubsection{Diagnosing Phase} 
As during sprint zero, all requirements related to eventual consistency had been dropped due to complexity, we conducted a focus group with the two full-stack developers who were already part of the development team during sprint zero. Our goal was to better understand the problem. Both members of the reference group also joined the focus group. The fourth author moderated this session, while the first author took notes and occasionally asked comprehension questions. For the first part of the focus group, the researchers prepared trigger questions to stimulate the discussion. These questions belonged to the following topic clusters: Consistency, Transactions, Asynchronous Update Propagation \& Conflict Management, Microservices, and DDD. The questions were also designed to test whether our anecdotal observations from past case studies \cite{DBLP:conf/eurosys/BraunBE21} could be accepted in the context of our study. In the second part, the attendants could freely share any thoughts regarding the topics discussed so far. We had planned the focus group to last 1.5 hours, but due to the very active participation of all attendants, we had to schedule a 30-minute follow-up meeting. The first researcher conducted a thematic analysis \cite{thematicAnalyisis} to analyze the transcripts of the focus group. The analysis results were later presented to the action team and the reference group and approved by all participants. 

The first researcher also carried out a baseline analysis on the existing model of the mood barometer app. To this end, she conducted a code review on the corresponding microservice repository and evaluated the quantitative metrics on the model in the code. She executed the evaluation of the metrics according to the design guidelines\footnote{The version of the guidelines used during the study is provided in the replication package:  https://doi.org/10.6084/m9.figshare.14988405}. 
The domain operations design guide provides instructions for assessing the compatibility of operations, whereas the domain objects design guide contains criteria for determining trivial aggregates. As the first researcher is the author of the guidelines, she had the expertise to correctly apply the metrics.

% \paragraph{Action Taking Phase}
\subsubsection{Action Taking Phase}
The first researcher prepared diary templates for the action team in each cycle. The action team members were asked to record if they observed anything good or bad in relation to the guidelines while working with them. Also, if action team members deliberately decided against the use of a guideline or best practice, they were asked to document this together with a  rationale. Action team members also used the diaries to share the relevant parts of the design, such as UML diagrams, signatures of domain operations, as well as domain logic in pseudo code listings. They were also asked to note for each aggregate its class according to our guidelines (see Section \ref{sec:domainObjectsDesign}). Similarly, regarding domain operations, they were asked to determine the update type, such as incremental update (see Section \ref{sec:domainOperationsDesign}), and the compatibility relations. 

In the first cycle, both developers independently developed a re-design proposal. In the second cycle, both developers collaboratively designed the model for upcoming user stories related to backlog grooming and effort estimations.

At the end of each action-taking phase, we conducted another focus group with the action team to collect and discuss feedback. Developers could freely share any feedback or observations, while the moderator posed questions aimed at challenging our applicability hypotheses given in Table \ref{tab:hypotheses}. 

% \paragraph{Evaluation Phase}
\subsubsection{Evaluation Phase}
During the evaluation phase, the first author reviewed the designs and determined the share of trivial aggregates and the share of compatible domain operations according to the guidelines. The evaluation of the metrics was conducted in the same way as during the evaluation of the baseline model, but this time using the model designs provided in the form of UML diagrams and pseudo code within the diaries. As the guidelines were revised in the action-taking phase of the second cycle, the evaluation of the metrics was adapted to the new guidelines during the second cycle. 

The first author also conducted a thematic analysis \cite{thematicAnalyisis} on qualitative data collected in diaries and the transcript of the focus group run at the end of each action-taking phase. In addition, she triangulated the qualitative data of the focus groups and the diaries with aggregate classification errors and domain operation compatibility assessment errors in order to substantiate the evidence of the derived learnings.

At the end of the evaluation phase, the researchers presented the results to the action team and the reference group. This meeting was also used to jointly conduct a focus group aimed at asking the reference group members for objective feedback. In the first cycle, we also used this focus group to diagnose the problems of the second cycle. Thus, the problems of the second cycle were diagnosed collaboratively by the action team and the reference group. 

% \vspace*{-0.5\baselineskip}
\section{Results and Interpretation}
\label{sec:results}
The results of the analysis of the quantitative data collected during both cycles\footnote{We provide the complete, manually anonimized dataset: https://doi.org/10.6084/m9.figshare.14988228} is given in Table \ref{tab:quantitativeResults}. Next, we report the results of the thematic analysis conducted on the qualitative data collected in the focus groups and diaries. The identified themes as well as code negatively or positively impacting our applicability hypotheses are presented in Tables \ref{tab:qualitativeResults1N}, \ref{tab:qualitativeResults1P} (cycle 1) and Tables \ref{tab:qualitativeResults2N}, \ref{tab:qualitativeResults2P} (cycle 2). 

The quantitative results given in Table \ref{tab:quantitativeResults} indicate that applying our design guidelines led to a high share of compatible operations (67\%-100\%). This was already the case in the first cycle, even though the qualitative instruments (see Table \ref{tab:qualitativeResults1N}) showed that the operation taxonomy on which our compatibility relation was based was hard to understand. Regarding the design of domain objects, trivial aggregate shares between 37.5\% and 60\% were achieved. 

\begin{table*}
    \caption{Quantitative Results Cycles 1 \& 2 - Effectiveness}
    \label{tab:quantitativeResults}
    \begin{minipage}{\textwidth}
    \begin{center}
    \begin{tabular}{p{0.18\linewidth} p{0.08\linewidth} p{0.3\linewidth} p{0.19\linewidth} p{0.16\linewidth}}
    \toprule
        \textbf{Bounded Context} & \textbf{Creation Time} & \textbf{Creator} & \textbf{Share of\newline Compatible \mbox{Operations}} & \textbf{Share of\newline Trivial \mbox{Aggregates}}\\
        \hline
        Team Mood Barometer & Baseline & D1 in collaboration with other developers & \phantom{10}\textbf{0.0\%} (0\phantom{1} out of \phantom{2}2) & \textbf{25.0\%} (1 out of 4)\\
        \hline
        Team Mood Barometer & Cycle 1 & D1 & \textbf{100.0\%} (4\phantom{1} out of \phantom{2}4) & \textbf{40.0\%} (2 out of 5)\\
        Team Mood Barometer & Cycle 1 & D2 & \phantom{1}\textbf{67.0\%} (2\phantom{1} out of \phantom{2}3) & \textbf{60.0\%} (3 out of 5)\\
        \hline
        Backlog Management & Cycle 2 & D1 in collaboration with D2 & \phantom{1}\textbf{95.5\%} (21 out of 22) & \textbf{37.5\%} (3 out of 8)\\
    \bottomrule
    \end{tabular}
    \end{center}
    \end{minipage}
\end{table*}

\begin{table*}
    \caption{Qualitative Results Cycle 1 - Codes with Negative Impact on Applicability}
    \label{tab:qualitativeResults1N}
    \begin{minipage}{\textwidth}
    \begin{center}
    \begin{tabular}{p{0.42\linewidth} p{0.55\linewidth}}
    \toprule
        \textbf{Theme \& Sub-Themes} & \textbf{Code with a Negative Impact}\\
        \hline
      H\textsubscript{App1} $\rightarrow$ Comprehensibility $\rightarrow$ \textbf{Aggregate Taxonomy} & Classification criteria are partly ambiguous\\
      & Some examples are ambiguous\\
      & Too little use of simple examples with which practitioners are already familiar\\
      & Some class names are counter-intuitive\\
      \hline
        H\textsubscript{App1} $\rightarrow$ Comprehensibility $\rightarrow$ \textbf{Operation Taxonomy} & Classification tree is hard to understand\\
        \hline
      H\textsubscript{App1} $\rightarrow$ Comprehensibility $\rightarrow$ \textbf{Compatibility Relations} & Compatibility relations not clear; require much more detailed explanations\\
      \hline
      H\textsubscript{App2} $\rightarrow$ \textbf{Applicability} & Re-design of an existing model in the context of a complex brownfield project\\
      & The existing model was complex and hard to understand\\
      & The existing model was an example of the anemic domain model anti-pattern\\
      & Unclear or incomplete business requirements\\
      & Missing cheat sheets and overview tables\\
      \hline
        H\textsubscript{App2} $\rightarrow$ Applicability $\rightarrow$ \textbf{Aggregate Taxonomy} & Classification criteria partly ambiguous\\
        \hline
        H\textsubscript{App2} $\rightarrow$ Applicability $\rightarrow$ \textbf{Operation Taxonomy} & Generic operations do not reflect the intent of users or the domain semantics\\
    \bottomrule
    \end{tabular}
    \end{center}
    \end{minipage}
\end{table*}

\begin{table*}
    \caption{Qualitative Results Cycle 1 - Codes with Positive Impact on Applicability}
    \label{tab:qualitativeResults1P}
    \begin{minipage}{\textwidth}
    \begin{center}
    \begin{tabular}{p{0.42\linewidth} p{0.55\linewidth}}
    \toprule
        \textbf{Theme \& Sub-Themes} & \textbf{Code with a Positive \mbox{Impact}}\\
        \hline
      H\textsubscript{App1} $\rightarrow$ Comprehensibility $\rightarrow$ \textbf{Aggregate Taxonomy} & Hands-on experience in using the taxonomy\\
      \hline
        H\textsubscript{App1} $\rightarrow$ Comprehensibility $\rightarrow$ \textbf{Operation Taxonomy} & Hands-on experience in using the taxonomy\\
      \hline
      H\textsubscript{App2} $\rightarrow$ \textbf{Applicability} & The re-designed  model was much clearer and easier to understand\\
      \hline
        H\textsubscript{App3} $\rightarrow$ \textbf{Complexity} &  After the re-design, the model was less complex \\
    \bottomrule
    \end{tabular}
    \end{center}
    \end{minipage}
\end{table*}

\begin{table*}
    \caption{Qualitative Results Cycle 2 - Codes with Negative Impact on Applicability}
    \label{tab:qualitativeResults2N}
    \begin{minipage}{\textwidth}
    \begin{center}
    \begin{tabular}{p{0.3\linewidth} p{0.67\linewidth}}
    \toprule
        \textbf{Theme \& Sub-Themes} & \textbf{Code with a Negative Impact}\\
        \hline
      H\textsubscript{App2} $\rightarrow$ Applicability $\rightarrow$ \textbf{Guidelines} & The corresponding programming framework is not available yet\\
      & Th exact features and guarantees of the corresponding programming framework are not clear\\
      & DDD novices -- missing hands-on experience in DDD\\
      & Superficial knowledge of central DDD concepts such as bounded contexts and aggregates\\
      & Substantial training effort required to learn central DDD concepts first\\
        & Missing concrete design patterns for certain aspects like the creation of new domain objects\\
        & Guidelines do not provide a recipe, but require understanding of the method\\
        & Unresolved questions regarding domain semantics\\
        & Too little direct communication between developers\\
      \hline
        H\textsubscript{App3} $\rightarrow$ \textbf{Complexity} & Segregation of different aggregate classes is sometimes perceived as unusual\\
    \bottomrule
    \end{tabular}
    \end{center}
    \end{minipage}
\end{table*}

\begin{table*}
    \caption{Qualitative Results Cycle 2 - Codes with Positive Impact on Applicability}
    \label{tab:qualitativeResults2P}
    \begin{minipage}{\textwidth}
    \begin{center}
    \begin{tabular}{p{0.35\linewidth} p{0.6\linewidth}}
    \toprule
        \textbf{Theme \& Sub-Themes} & \textbf{Code with a Positive \mbox{Impact}}\\
        \hline
      H\textsubscript{App1} $\rightarrow$ Comprehensibility $\rightarrow$ \textbf{Guidelines} & Guidelines are understandable\\
      \hline
      H\textsubscript{App2} $\rightarrow$ Applicability $\rightarrow$ \textbf{Guidelines} & Guidelines are applicable \\
      &  Examples are transferable \\
      \hline
        H\textsubscript{App3} $\rightarrow$ \textbf{Complexity} & Perceived additional complexity is low \\
    \bottomrule
    \end{tabular}
    \end{center}
    \end{minipage}
\end{table*}

% \vspace*{-\baselineskip}
\subsection{Cycle 1}
% \paragraph{Cycle 1}
The share of trivial aggregates in the re-designs of the developers was between 40\% and 60\%. The potential to design parts of the model with trivial aggregates is highly dependent on the domain semantics. However, the first author, who also conducted the assessment of the designs, was able to come up with a team mood barometer design with 4 out of 5 aggregates being trivial. Thus, in principle, a share of 80\% would have been possible. On the downside, the model of the researcher contained a notable, but acceptable share of code duplication. One example where this can be seen are aggregates that, depending on their lifecycle, need to reference particular versions of other aggregates. For instance, each team mood barometer has a configuration. Multiple configurations can be maintained by one team and can be adapted to meet changing needs of the team. But once a team mood barometer has gone live and team members can place votes, the configuration must not be changed anymore.
The researcher did not reuse mutable configuration objects to model the relation between a mood barometer gone live and its configuration. Instead, she modeled an explicit immutable configuration aggregate that did not offer any update operations in its interface. She made the immutability property very explicit in the model, but had to accept a certain amount of code duplication in return. In contrast, the developers seemed to have a stronger tendency towards generalization and reuse of existing model elements. This was especially true for the baseline model and also explains why the number of aggregates was smaller in the baseline model compared to the re-designs. 
The tendency to generalize led to an increase of complexity, as, e.g., the developers had to come up with the correct logic to decide whether or not the configuration aggregate was in the state to accept updates 
(which they did not do correctly in the baseline code). 

The code review of the baseline model showed that the baseline model 
% Regarding the design of domain operations in the baseline model, the code review showed that the model
was an example of the anemic domain model anti-pattern \cite{FowlerAnemicDomainModel} with only two coarse-grained state-based update operations for two aggregates. These operations would have been prone to lost updates if propagated in a setting with eventually consistent replication. However, to be fair, the developers originally did not create the model to be safe with eventually consistent replication as corresponding requirements had been dropped during sprint zero. Thus, we cannot exclude for sure that they would have been able to come up with a safe design without our guidelines. We did, however, directly ask this question during the last focus group of the second cycle and the developers denied it. 

One of the positive outcomes of the qualitative instruments was that the developers perceived the re-designed model as clearer and less complex. One developer directly stated that by applying the guidelines, things became more explicit in the model and that the guidelines forced them to be more explicit. The developers perceived this as helpful for structuring the model and for coming up with an improved design. This gain may also be due in part to the enforcement of DDD itself (as a basis of our guidelines). Therefore, we explicitly state this as a potential confounding factor.

The analysis of the designs combined with the results of the thematic analysis showed that the guidelines needed improvement in several areas. Comprehensibility and applicability in general were impaired, among other things, by ambiguous examples and classification criteria, or by counter-intuitive names of classes and compatibility relations. In particular, the guidelines for compatible domain operations design needed a major revision, as the developers were somewhat able to apply them but did not really understand them. Other factors the developers perceived as negatively impacting applicability are given in Table \ref{tab:qualitativeResults1N}. 

During the final focus group, the researchers also got the impression that one developer had been afraid to fundamentally re-design the baseline model because of the expected high refactoring effort. We interpreted this as a possible reason for not ``freely'' applying all our guidelines. The second developer was biased towards the baseline model because he had originally created it. Therefore, the researchers suggested challenging the revised guidelines during the second cycle on a model that could be designed from scratch.

% \vspace*{-\baselineskip}
\subsection{Cycle 2}
% \paragraph{Cycle 2}
% Interpretation of the results of the 2nd cycle.
In the second cycle, the developers achieved a very high share of compatible domain operations (95.5\%). Only one out of 22 operations was incompatible. The developers designed this operation with a blind update that is, in fact, incremental: collaborative text editing of user stories. However, collaborative text editing is challenging. Nevertheless, a possible solution would have been to search for a collaborative text-editing library based on CRDTs \cite{DBLP:conf/sss/ShapiroPBZ11} (e.g., \cite{yjs, Concordant}). The developers knew CRDTs as the researchers had presented CRDTs during the action-taking phase of the first cycle. The share of trivial aggregates was 37.5\%, which could not have been increased meaningfully. The potential for trivial aggregates was simply not so high in the backlog management model as in the model of cycle 1. 

Note that the size of the domain model of the second cycle was significantly larger in terms of aggregates count and number of domain operations. Hence, the action team decided to design only one model in collaboration with both developers. This way of working also represented our usual mode of operation in the project. Thus, in the second cycle, we traded off larger amounts of quantitative data for increased realism.

The thematic analysis revealed that the developers perceived the overall additional complexity introduced by following the recommendations of the guidelines as low. Furthermore, they clearly stated that the revised guidelines were supportive, comprehensible, and applicable. The developers also praised the introduction of additional cheat sheets. However, the participants would have wished for a programming framework directly supporting all the technical prerequisites outlined in the guidelines. With regard to the domain operations design guidelines, the participants mentioned that the possibility to prototype certain aspects of the model directly in code would have been beneficial. Another weakness was that applying the guidelines already requires an advanced level of experience with DDD. The developers admitted that they considered themselves rather to be DDD novices, and therefore their overall learning curve had been quite steep. In particular, they declared that they had never before applied the DDD aggregate pattern in practice. They also noted that regarding trivial aggregates, they sometimes perceived it as unusual to segregate different aggregate classes (segregation of different aggregate classes in the model is one of the best practices from our guidelines).

% \raggedbottom
\section{Validity}
\label{sec:validity}
% 0,5 Pages
Each empirical study comes with threats to validity. We discuss the most critical ones based on the guidelines by Staron \cite{DBLP:books/sp/Staron20}.

\textit{Construct validity – mono operation bias}: In our study, the action team consisted of four people. The two action research cycles during which the tasks were performed ran from October 2020 to March 2021.

\textit{Construct validity – hypotheses guessing}: As the action team tested guidelines to get results, a certain expectation regarding improved results cannot be neglected. We explicitly asked whether the action team had produced the results to meet possible result expectations, which they denied. Finally, we discussed not only improvements, but also critical aspects regarding our guidelines and results. 

\textit{Construct validity – evaluation apprehension}: We decided against direct observation, but asked the action team to use a diary and provide feedback in a meeting.

\textit{Construct validity – experimenter expectations}: A positive expectation regarding the results cannot be completely denied, and the concrete tasks were selected so that the guidelines could be applied. However, the setting of the study was not specifically adapted, and we discussed the tasks with the reference group and the second researcher, who mainly helped to organize the action research cycles. As he was not deeply involved in content creation, he was less biased.

\textit{Internal validity – maturation}: Over the course of the two cycles, a certain maturation of the participants cannot be completely neglected. But the concrete tasks were different in the two cycles. Furthermore, the timeframe for using the guidelines in the concrete tasks was short (about 1-2 weeks each) so that maturation did not appear to have a strong influencing effect.

\textit{Conclusion validity – changed procedures of action team}: We changed the procedure between the two action research cycles. In the first cycle, every person solved the tasks on their own. In the second cycle, a two-person team was defined. This reflects the real-world character better, but means that the direct comparison of the two runs had to be conducted carefully.

\textit{External validity – real-world setting vs. experimental setting}: The results are part of a publicly funded research project. However, the background is strongly related to industry, and the concrete project is a realistic example from a running system relevant for industrial settings. 
% \vspace*{-\baselineskip}
\section{Summary \& Conclusions}\label{sec:summmary-conclusions}
In our action research, we set off to find new ways for tackling consistency-related design challenges of distributed data-intensive systems. We applied a novel method for designing safe eventually consistent domain models. The following key results were obtained: (1) a high share of compatible domain operations in the resulting models; (2) high exploitation of the potential to design parts of the model with trivial aggregates; (3) the developers' perception of the guidelines being supportive, comprehensible, and applicable; (4) the developers' declaration that they would not have been able to consider all aspects covered by our design guidelines on their own; (5) the developers' perception of the resulting domain models being easier to understand and less complex in comparison to the baseline; (6) the developers' perception of additional complexity introduced by the application of our guidelines being low. 

Thus, our initial assumptions were confirmed and we can positively answer our research question (RQ). We accept our research hypotheses concerning the effectiveness and applicability of our design guidelines in the context of our study. Based on our results, we recommend using the guidelines to design safe eventually consistent domain models, even though the current version lacks support, e.g., by a programming framework, and requires an advanced level of experience with DDD.

To obtain more empirical evidence about the applicability and efficiency of our approach, we regularly conduct hands-on, half-day workshops with external practitioners, teaching them our design method and collecting feedback. Furthermore, we will develop and evaluate a programming framework integrated with Enterprise Java\footnote{https://www.java.com/de/} technology to support developers with regard to the technical prerequisites outlined in the guidelines.

\begin{acks}
Many thanks to the members of the action team and the reference group. We thank Sonnhild Namingha for proofreading and our anonymous reviewers for valuable feedback. The action research study has been
partially funded by the German Federal Ministry of Economic Affairs and
Energy (01MD18007C).
\end{acks}

\bibliographystyle{ACM-Reference-Format}
\bibliography{references}


\begin{thebibliography}{47}

%%% the Web address.

\ifx \showCODEN    \undefined \def \showCODEN     #1{\unskip}     \fi
\ifx \showDOI      \undefined \def \showDOI       #1{#1}\fi
\ifx \showISBNx    \undefined \def \showISBNx     #1{\unskip}     \fi
\ifx \showISBNxiii \undefined \def \showISBNxiii  #1{\unskip}     \fi
\ifx \showISSN     \undefined \def \showISSN      #1{\unskip}     \fi
\ifx \showLCCN     \undefined \def \showLCCN      #1{\unskip}     \fi
\ifx \shownote     \undefined \def \shownote      #1{#1}          \fi
\ifx \showarticletitle \undefined \def \showarticletitle #1{#1}   \fi
\ifx \showURL      \undefined \def \showURL       {\relax}        \fi
% invisible to TeX
\providecommand\bibfield[2]{#2}
\providecommand\bibinfo[2]{#2}
\providecommand\natexlab[1]{#1}
\providecommand\showeprint[2][]{arXiv:#2}

\bibitem[\protect\citeauthoryear{??}{Con}{[n.d.]}]%
        {Concordant}
 \bibinfo{year}{[n.d.]}\natexlab{}.
\newblock \bibinfo{booktitle}{\emph{Concordant}}.
\newblock
\urldef\tempurl%
\url{https://concordant.io/}
\showURL{%
Retrieved Mai 03, 2021 from \tempurl}


\bibitem[\protect\citeauthoryear{Ahmed and Bagherzadeh}{Ahmed and
  Bagherzadeh}{2018}]%
        {DBLP:conf/esem/AhmedB18}
\bibfield{author}{\bibinfo{person}{Syed Ahmed} {and} \bibinfo{person}{Mehdi
  Bagherzadeh}.} \bibinfo{year}{2018}\natexlab{}.
\newblock \showarticletitle{What do concurrency developers ask about?: a
  large-scale study using stack overflow}. In
  \bibinfo{booktitle}{\emph{Proceedings of the 12th {ACM/IEEE} International
  Symposium on Empirical Software Engineering and Measurement, {ESEM} 2018,
  Oulu, Finland, October 11-12, 2018}},
  \bibfield{editor}{\bibinfo{person}{Markku Oivo},
  \bibinfo{person}{Daniel~M{\'{e}}ndez Fern{\'{a}}ndez}, {and}
  \bibinfo{person}{Audris Mockus}} (Eds.). \bibinfo{publisher}{{ACM}},
  \bibinfo{pages}{30:1--30:10}.
\newblock
\urldef\tempurl%
\url{https://doi.org/10.1145/3239235.3239524}
\showDOI{\tempurl}


\bibitem[\protect\citeauthoryear{Berenson, Bernstein, Gray, Melton, O'Neil, and
  O'Neil}{Berenson et~al\mbox{.}}{1995}]%
        {DBLP:conf/sigmod/BerensonBGMOO95}
\bibfield{author}{\bibinfo{person}{Hal Berenson}, \bibinfo{person}{Philip~A.
  Bernstein}, \bibinfo{person}{Jim Gray}, \bibinfo{person}{Jim Melton},
  \bibinfo{person}{Elizabeth~J. O'Neil}, {and} \bibinfo{person}{Patrick~E.
  O'Neil}.} \bibinfo{year}{1995}\natexlab{}.
\newblock \showarticletitle{A Critique of {ANSI} {SQL} Isolation Levels}. In
  \bibinfo{booktitle}{\emph{Proceedings of the 1995 {ACM} {SIGMOD}
  International Conference on Management of Data, San Jose, California, USA,
  May 22-25, 1995}}, \bibfield{editor}{\bibinfo{person}{Michael~J. Carey} {and}
  \bibinfo{person}{Donovan~A. Schneider}} (Eds.). \bibinfo{publisher}{{ACM}
  Press}, \bibinfo{pages}{1--10}.
\newblock
\showISBNx{978-0-89791-731-5}
\urldef\tempurl%
\url{https://doi.org/10.1145/223784.223785}
\showDOI{\tempurl}


\bibitem[\protect\citeauthoryear{Braun}{Braun}{2017}]%
        {DBLP:conf/icsa/Braun17}
\bibfield{author}{\bibinfo{person}{Susanne Braun}.}
  \bibinfo{year}{2017}\natexlab{}.
\newblock \showarticletitle{Semantics-Driven Optimistic Data Replication:
  Towards a Framework Supporting Software Architects and Developers}. In
  \bibinfo{booktitle}{\emph{2017 {IEEE} International Conference on Software
  Architecture Workshops, {ICSA} Workshops 2017, Gothenburg, Sweden, April 5-7,
  2017}}. \bibinfo{publisher}{{IEEE} Computer Society},
  \bibinfo{pages}{236--241}.
\newblock
\showISBNx{978-1-5090-4793-2}
\urldef\tempurl%
\url{https://doi.org/10.1109/ICSAW.2017.18}
\showDOI{\tempurl}


\bibitem[\protect\citeauthoryear{Braun, Bieniusa, and Elberzhager}{Braun
  et~al\mbox{.}}{2021}]%
        {DBLP:conf/eurosys/BraunBE21}
\bibfield{author}{\bibinfo{person}{Susanne Braun}, \bibinfo{person}{Annette
  Bieniusa}, {and} \bibinfo{person}{Frank Elberzhager}.}
  \bibinfo{year}{2021}\natexlab{}.
\newblock \showarticletitle{Advanced Domain-Driven Design for Consistency in
  Distributed Data-Intensive Systems}. In
  \bibinfo{booktitle}{\emph{PaPoC@EuroSys 2021, 8th Workshop on Principles and
  Practice of Consistency for Distributed Data, Online Event, United Kingdom,
  April 26, 2021}}. \bibinfo{publisher}{{ACM}}, \bibinfo{pages}{9:1--9:12}.
\newblock
\urldef\tempurl%
\url{https://doi.org/10.1145/3447865.3457969}
\showDOI{\tempurl}


\bibitem[\protect\citeauthoryear{Braun, Carbon, and Naab}{Braun
  et~al\mbox{.}}{2016}]%
        {DBLP:journals/software/BraunCN16}
\bibfield{author}{\bibinfo{person}{Susanne Braun}, \bibinfo{person}{Ralf
  Carbon}, {and} \bibinfo{person}{Matthias Naab}.}
  \bibinfo{year}{2016}\natexlab{}.
\newblock \showarticletitle{Piloting a Mobile-App Ecosystem for Smart Farming}.
\newblock \bibinfo{journal}{\emph{{IEEE} Softw.}} \bibinfo{volume}{33},
  \bibinfo{number}{4} (\bibinfo{year}{2016}), \bibinfo{pages}{9--14}.
\newblock
\urldef\tempurl%
\url{https://doi.org/10.1109/MS.2016.98}
\showDOI{\tempurl}


\bibitem[\protect\citeauthoryear{Braun and De{\ss}loch}{Braun and
  De{\ss}loch}{2020}]%
        {DBLP:conf/icsa/BraunD20}
\bibfield{author}{\bibinfo{person}{Susanne Braun} {and} \bibinfo{person}{Stefan
  De{\ss}loch}.} \bibinfo{year}{2020}\natexlab{}.
\newblock \showarticletitle{A Classification of Replicated Data for the Design
  of Eventually Consistent Domain Models}. In \bibinfo{booktitle}{\emph{2020
  {IEEE} International Conference on Software Architecture Companion, {ICSA}
  Companion 2020, Salvador, Brazil, March 16-20, 2020}}.
  \bibinfo{publisher}{{IEEE}}, \bibinfo{pages}{33--40}.
\newblock
\showISBNx{978-1-7281-7415-0}
\urldef\tempurl%
\url{https://doi.org/10.1109/ICSA-C50368.2020.00014}
\showDOI{\tempurl}


\bibitem[\protect\citeauthoryear{Braun and Clarke}{Braun and Clarke}{2006}]%
        {thematicAnalyisis}
\bibfield{author}{\bibinfo{person}{Virginia Braun} {and}
  \bibinfo{person}{Victoria Clarke}.} \bibinfo{year}{2006}\natexlab{}.
\newblock \showarticletitle{Using thematic analysis in psychology}.
\newblock \bibinfo{journal}{\emph{Qualitative research in psychology}}
  \bibinfo{volume}{3}, \bibinfo{number}{2} (\bibinfo{year}{2006}),
  \bibinfo{pages}{77--101}.
\newblock


\bibitem[\protect\citeauthoryear{Breitbart, Garcia{-}Molina, and
  Silberschatz}{Breitbart et~al\mbox{.}}{1992}]%
        {DBLP:journals/vldb/BreitbartGS92}
\bibfield{author}{\bibinfo{person}{Yuri Breitbart}, \bibinfo{person}{Hector
  Garcia{-}Molina}, {and} \bibinfo{person}{Abraham Silberschatz}.}
  \bibinfo{year}{1992}\natexlab{}.
\newblock \showarticletitle{Overview of Multidatabase Transaction Management}.
\newblock \bibinfo{journal}{\emph{{VLDB} J.}} \bibinfo{volume}{1},
  \bibinfo{number}{2} (\bibinfo{year}{1992}), \bibinfo{pages}{181--239}.
\newblock
\urldef\tempurl%
\url{http://www.vldb.org/journal/VLDBJ1/P181.pdf}
\showURL{%
\tempurl}


\bibitem[\protect\citeauthoryear{Brewer}{Brewer}{2012}]%
        {brewer2012cap}
\bibfield{author}{\bibinfo{person}{Eric Brewer}.}
  \bibinfo{year}{2012}\natexlab{}.
\newblock \showarticletitle{CAP twelve years later: How the" rules" have
  changed}.
\newblock \bibinfo{journal}{\emph{Computer}} \bibinfo{volume}{45},
  \bibinfo{number}{2} (\bibinfo{year}{2012}), \bibinfo{pages}{23--29}.
\newblock


\bibitem[\protect\citeauthoryear{Brewer}{Brewer}{2017}]%
        {brewer2017spanner}
\bibfield{author}{\bibinfo{person}{Eric Brewer}.}
  \bibinfo{year}{2017}\natexlab{}.
\newblock \showarticletitle{Spanner, truetime and the cap theorem}.
\newblock  (\bibinfo{year}{2017}).
\newblock


\bibitem[\protect\citeauthoryear{Brewer}{Brewer}{2000}]%
        {DBLP:conf/podc/Brewer00}
\bibfield{author}{\bibinfo{person}{Eric~A. Brewer}.}
  \bibinfo{year}{2000}\natexlab{}.
\newblock \showarticletitle{Towards robust distributed systems (abstract)}. In
  \bibinfo{booktitle}{\emph{Proceedings of the Nineteenth Annual {ACM}
  Symposium on Principles of Distributed Computing, July 16-19, 2000, Portland,
  Oregon, {USA}}}, \bibfield{editor}{\bibinfo{person}{Gil Neiger}} (Ed.).
  \bibinfo{publisher}{{ACM}}, \bibinfo{pages}{7}.
\newblock
\showISBNx{1-58113-183-6}
\urldef\tempurl%
\url{https://doi.org/10.1145/343477.343502}
\showDOI{\tempurl}


\bibitem[\protect\citeauthoryear{Caldiera and Rombach}{Caldiera and
  Rombach}{1994}]%
        {caldiera1994goal}
\bibfield{author}{\bibinfo{person}{Victor R Basili-Gianluigi Caldiera} {and}
  \bibinfo{person}{H~Dieter Rombach}.} \bibinfo{year}{1994}\natexlab{}.
\newblock \showarticletitle{Goal question metric paradigm}.
\newblock \bibinfo{journal}{\emph{Encyclopedia of software engineering}}
  \bibinfo{volume}{1}, \bibinfo{number}{528-532} (\bibinfo{year}{1994}),
  \bibinfo{pages}{6}.
\newblock


\bibitem[\protect\citeauthoryear{Corbett, Dean, Epstein, Fikes, Frost, Furman,
  Ghemawat, Gubarev, Heiser, Hochschild, et~al\mbox{.}}{Corbett
  et~al\mbox{.}}{2013}]%
        {corbett2013spanner}
\bibfield{author}{\bibinfo{person}{James~C Corbett}, \bibinfo{person}{Jeffrey
  Dean}, \bibinfo{person}{Michael Epstein}, \bibinfo{person}{Andrew Fikes},
  \bibinfo{person}{Christopher Frost}, \bibinfo{person}{Jeffrey~John Furman},
  \bibinfo{person}{Sanjay Ghemawat}, \bibinfo{person}{Andrey Gubarev},
  \bibinfo{person}{Christopher Heiser}, \bibinfo{person}{Peter Hochschild},
  {et~al\mbox{.}}} \bibinfo{year}{2013}\natexlab{}.
\newblock \showarticletitle{Spanner: Google’s globally distributed database}.
\newblock \bibinfo{journal}{\emph{ACM Transactions on Computer Systems (TOCS)}}
  \bibinfo{volume}{31}, \bibinfo{number}{3} (\bibinfo{year}{2013}),
  \bibinfo{pages}{1--22}.
\newblock


\bibitem[\protect\citeauthoryear{Elmagarmid}{Elmagarmid}{1992}]%
        {DBLP:books/mk/Elmagarmid92}
\bibfield{editor}{\bibinfo{person}{Ahmed~K. Elmagarmid}} (Ed.).
  \bibinfo{year}{1992}\natexlab{}.
\newblock \bibinfo{booktitle}{\emph{Database Transaction Models for Advanced
  Applications}}.
\newblock In Elmagarmid \citeN{DBLP:books/mk/Elmagarmid92}.
\newblock
\showISBNx{1-55860-214-3}


\bibitem[\protect\citeauthoryear{Evans}{Evans}{2004}]%
        {DBLP:books/EvansDDD}
\bibfield{author}{\bibinfo{person}{Eric~J. Evans}.}
  \bibinfo{year}{2004}\natexlab{}.
\newblock \bibinfo{booktitle}{\emph{Domain-driven design - tackling complexity
  in the heart of software}}.
\newblock \bibinfo{publisher}{Addison-Wesley}.
\newblock
\showISBNx{978-0-321-12521-7}


\bibitem[\protect\citeauthoryear{Falessi, Babar, Cantone, and Kruchten}{Falessi
  et~al\mbox{.}}{2010}]%
        {DBLP:journals/ese/FalessiBCK10}
\bibfield{author}{\bibinfo{person}{Davide Falessi},
  \bibinfo{person}{Muhammad~Ali Babar}, \bibinfo{person}{Giovanni Cantone},
  {and} \bibinfo{person}{Philippe Kruchten}.} \bibinfo{year}{2010}\natexlab{}.
\newblock \showarticletitle{Applying empirical software engineering to software
  architecture: challenges and lessons learned}.
\newblock \bibinfo{journal}{\emph{Empir. Softw. Eng.}} \bibinfo{volume}{15},
  \bibinfo{number}{3} (\bibinfo{year}{2010}), \bibinfo{pages}{250--276}.
\newblock
\urldef\tempurl%
\url{https://doi.org/10.1007/s10664-009-9121-0}
\showDOI{\tempurl}


\bibitem[\protect\citeauthoryear{Fowler}{Fowler}{2003}]%
        {FowlerAnemicDomainModel}
\bibfield{author}{\bibinfo{person}{Martin Fowler}.}
  \bibinfo{year}{2003}\natexlab{}.
\newblock \bibinfo{booktitle}{\emph{Anemic Domain Model}}.
\newblock
\urldef\tempurl%
\url{https://www.martinfowler.com/bliki/AnemicDomainModel.html}
\showURL{%
Retrieved March 25, 2021 from \tempurl}


\bibitem[\protect\citeauthoryear{Fowler}{Fowler}{2005}]%
        {FowlerEventSourcing}
\bibfield{author}{\bibinfo{person}{Martin Fowler}.}
  \bibinfo{year}{2005}\natexlab{}.
\newblock \bibinfo{booktitle}{\emph{Event Sourcing}}.
\newblock
\urldef\tempurl%
\url{https://martinfowler.com/eaaDev/EventSourcing.html}
\showURL{%
Retrieved March 25, 2021 from \tempurl}


\bibitem[\protect\citeauthoryear{Fowler}{Fowler}{2011}]%
        {cqrs}
\bibfield{author}{\bibinfo{person}{Martin Fowler}.}
  \bibinfo{year}{2011}\natexlab{}.
\newblock \bibinfo{booktitle}{\emph{CQRS}}.
\newblock
\urldef\tempurl%
\url{https://martinfowler.com/bliki/CQRS.html}
\showURL{%
Retrieved May 24, 2020 from \tempurl}


\bibitem[\protect\citeauthoryear{Fowler}{Fowler}{2015}]%
        {FowlerMicroserviceTradeoffs}
\bibfield{author}{\bibinfo{person}{Martin Fowler}.}
  \bibinfo{year}{2015}\natexlab{}.
\newblock \bibinfo{booktitle}{\emph{Microservice Trade-Offs}}.
\newblock
\urldef\tempurl%
\url{https://martinfowler.com/articles/microservice-trade-offs.html#consistency}
\showURL{%
Retrieved Feb 05, 2021 from \tempurl}


\bibitem[\protect\citeauthoryear{Garcia{-}Molina and Salem}{Garcia{-}Molina and
  Salem}{1987}]%
        {DBLP:conf/sigmod/Garcia-MolinaS87}
\bibfield{author}{\bibinfo{person}{Hector Garcia{-}Molina} {and}
  \bibinfo{person}{Kenneth Salem}.} \bibinfo{year}{1987}\natexlab{}.
\newblock \showarticletitle{Sagas}. In \bibinfo{booktitle}{\emph{Proceedings of
  the Association for Computing Machinery Special Interest Group on Management
  of Data 1987 Annual Conference, San Francisco, CA, USA, May 27-29, 1987}},
  \bibfield{editor}{\bibinfo{person}{Umeshwar Dayal} {and}
  \bibinfo{person}{Irving~L. Traiger}} (Eds.). \bibinfo{publisher}{{ACM}
  Press}, \bibinfo{pages}{249--259}.
\newblock
\showISBNx{978-0-89791-236-5}
\urldef\tempurl%
\url{https://doi.org/10.1145/38713.38742}
\showDOI{\tempurl}


\bibitem[\protect\citeauthoryear{Godefroid and Nagappan}{Godefroid and
  Nagappan}{2008}]%
        {godefroid2008concurrency}
\bibfield{author}{\bibinfo{person}{Patrice Godefroid} {and}
  \bibinfo{person}{Nachiappan Nagappan}.} \bibinfo{year}{2008}\natexlab{}.
\newblock \showarticletitle{Concurrency at Microsoft: An exploratory survey}.
  In \bibinfo{booktitle}{\emph{CAV workshop on exploiting concurrency
  efficiently and correctly}}. Princeton, USA.
\newblock


\bibitem[\protect\citeauthoryear{Gray, Helland, O'Neil, and Shasha}{Gray
  et~al\mbox{.}}{1996}]%
        {DBLP:conf/sigmod/GrayHOS96}
\bibfield{author}{\bibinfo{person}{Jim Gray}, \bibinfo{person}{Pat Helland},
  \bibinfo{person}{Patrick~E. O'Neil}, {and} \bibinfo{person}{Dennis~E.
  Shasha}.} \bibinfo{year}{1996}\natexlab{}.
\newblock \showarticletitle{The Dangers of Replication and a Solution}. In
  \bibinfo{booktitle}{\emph{Proceedings of the 1996 {ACM} {SIGMOD}
  International Conference on Management of Data, Montreal, Quebec, Canada,
  June 4-6, 1996}}, \bibfield{editor}{\bibinfo{person}{H.~V. Jagadish} {and}
  \bibinfo{person}{Inderpal~Singh Mumick}} (Eds.). \bibinfo{publisher}{{ACM}
  Press}, \bibinfo{pages}{173--182}.
\newblock
\showISBNx{978-0-89791-794-0}
\urldef\tempurl%
\url{https://doi.org/10.1145/233269.233330}
\showDOI{\tempurl}


\bibitem[\protect\citeauthoryear{Gray and Reuter}{Gray and Reuter}{1993}]%
        {DBLP:books/mk/GrayR93}
\bibfield{author}{\bibinfo{person}{Jim Gray} {and} \bibinfo{person}{Andreas
  Reuter}.} \bibinfo{year}{1993}\natexlab{}.
\newblock \bibinfo{booktitle}{\emph{Transaction Processing: Concepts and
  Techniques}}.
\newblock \bibinfo{publisher}{Morgan Kaufmann}.
\newblock
\showISBNx{1-55860-190-2}


\bibitem[\protect\citeauthoryear{H{\"{a}}rder and Reuter}{H{\"{a}}rder and
  Reuter}{1983}]%
        {DBLP:journals/csur/HaerderR83}
\bibfield{author}{\bibinfo{person}{Theo H{\"{a}}rder} {and}
  \bibinfo{person}{Andreas Reuter}.} \bibinfo{year}{1983}\natexlab{}.
\newblock \showarticletitle{Principles of Transaction-Oriented Database
  Recovery}.
\newblock \bibinfo{journal}{\emph{{ACM} Comput. Surv.}} \bibinfo{volume}{15},
  \bibinfo{number}{4} (\bibinfo{year}{1983}), \bibinfo{pages}{287--317}.
\newblock
\urldef\tempurl%
\url{https://doi.org/10.1145/289.291}
\showDOI{\tempurl}


\bibitem[\protect\citeauthoryear{Hasselbring and Steinacker}{Hasselbring and
  Steinacker}{2017}]%
        {DBLP:conf/icsa/HasselbringS17}
\bibfield{author}{\bibinfo{person}{Wilhelm Hasselbring} {and}
  \bibinfo{person}{Guido Steinacker}.} \bibinfo{year}{2017}\natexlab{}.
\newblock \showarticletitle{Microservice Architectures for Scalability, Agility
  and Reliability in E-Commerce}. In \bibinfo{booktitle}{\emph{2017 {IEEE}
  International Conference on Software Architecture Workshops, {ICSA} Workshops
  2017, Gothenburg, Sweden, April 5-7, 2017}}. \bibinfo{publisher}{{IEEE}
  Computer Society}, \bibinfo{pages}{243--246}.
\newblock
\urldef\tempurl%
\url{https://doi.org/10.1109/ICSAW.2017.11}
\showDOI{\tempurl}


\bibitem[\protect\citeauthoryear{Helland and Campbell}{Helland and
  Campbell}{2009}]%
        {DBLP:conf/cidr/HellandC09}
\bibfield{author}{\bibinfo{person}{Pat Helland} {and} \bibinfo{person}{David
  Campbell}.} \bibinfo{year}{2009}\natexlab{}.
\newblock \showarticletitle{Building on Quicksand}. In
  \bibinfo{booktitle}{\emph{{CIDR} 2009, Fourth Biennial Conference on
  Innovative Data Systems Research, Asilomar, CA, USA, January 4-7, 2009,
  Online Proceedings}}. \bibinfo{publisher}{www.cidrdb.org}.
\newblock
\urldef\tempurl%
\url{http://www-db.cs.wisc.edu/cidr/cidr2009/Paper\_133.pdf}
\showURL{%
\tempurl}


\bibitem[\protect\citeauthoryear{Hellerstein and Alvaro}{Hellerstein and
  Alvaro}{2020}]%
        {DBLP:journals/cacm/HellersteinA20}
\bibfield{author}{\bibinfo{person}{Joseph~M. Hellerstein} {and}
  \bibinfo{person}{Peter Alvaro}.} \bibinfo{year}{2020}\natexlab{}.
\newblock \showarticletitle{Keeping {CALM:} when distributed consistency is
  easy}.
\newblock \bibinfo{journal}{\emph{Commun. {ACM}}} \bibinfo{volume}{63},
  \bibinfo{number}{9} (\bibinfo{year}{2020}), \bibinfo{pages}{72--81}.
\newblock
\urldef\tempurl%
\url{https://dl.acm.org/doi/10.1145/3369736}
\showURL{%
\tempurl}


\bibitem[\protect\citeauthoryear{Jahns}{Jahns}{[n.d.]}]%
        {yjs}
\bibfield{author}{\bibinfo{person}{Kevin Jahns}.}
  \bibinfo{year}{[n.d.]}\natexlab{}.
\newblock \bibinfo{booktitle}{\emph{yjs}}.
\newblock
\urldef\tempurl%
\url{https://github.com/yjs/yjs}
\showURL{%
Retrieved Feb 06, 2021 from \tempurl}


\bibitem[\protect\citeauthoryear{Kleppmann}{Kleppmann}{2016}]%
        {DBLP:books/oreilly/Kleppmann2014}
\bibfield{author}{\bibinfo{person}{Martin Kleppmann}.}
  \bibinfo{year}{2016}\natexlab{}.
\newblock \bibinfo{booktitle}{\emph{Designing Data-Intensive Applications: The
  Big Ideas Behind Reliable, Scalable, and Maintainable Systems}}.
\newblock \bibinfo{publisher}{O'Reilly}.
\newblock
\showISBNx{978-1-4493-7332-0}
\urldef\tempurl%
\url{http://shop.oreilly.com/product/0636920032175.do}
\showURL{%
\tempurl}


\bibitem[\protect\citeauthoryear{Naab, Braun, Lenhart, Hess, Eitel, Magin,
  Carbon, and Kiefer}{Naab et~al\mbox{.}}{2015}]%
        {DBLP:conf/wicsa/NaabBLHEMCK15}
\bibfield{author}{\bibinfo{person}{Matthias Naab}, \bibinfo{person}{Susanne
  Braun}, \bibinfo{person}{Torsten Lenhart}, \bibinfo{person}{Steffen Hess},
  \bibinfo{person}{Andreas Eitel}, \bibinfo{person}{Dominik Magin},
  \bibinfo{person}{Ralf Carbon}, {and} \bibinfo{person}{Felix Kiefer}.}
  \bibinfo{year}{2015}\natexlab{}.
\newblock \showarticletitle{Why Data Needs more Attention in Architecture
  Design - Experiences from Prototyping a Large-Scale Mobile App Ecosystem}. In
  \bibinfo{booktitle}{\emph{12th Working {IEEE/IFIP} Conference on Software
  Architecture, {WICSA} 2015, Montreal, QC, Canada, May 4-8, 2015}},
  \bibfield{editor}{\bibinfo{person}{Len Bass}, \bibinfo{person}{Patricia
  Lago}, {and} \bibinfo{person}{Philippe Kruchten}} (Eds.).
  \bibinfo{publisher}{{IEEE} Computer Society}, \bibinfo{pages}{75--84}.
\newblock
\showISBNx{978-1-4799-1922-2}
\urldef\tempurl%
\url{https://doi.org/10.1109/WICSA.2015.13}
\showDOI{\tempurl}


\bibitem[\protect\citeauthoryear{Pardon, Pautasso, and Zimmermann}{Pardon
  et~al\mbox{.}}{2018}]%
        {DBLP:journals/cloudcomp/PardonPZ18}
\bibfield{author}{\bibinfo{person}{Guy Pardon}, \bibinfo{person}{Cesare
  Pautasso}, {and} \bibinfo{person}{Olaf Zimmermann}.}
  \bibinfo{year}{2018}\natexlab{}.
\newblock \showarticletitle{Consistent Disaster Recovery for Microservices: the
  {BAC} Theorem}.
\newblock \bibinfo{journal}{\emph{{IEEE} Cloud Comput.}} \bibinfo{volume}{5},
  \bibinfo{number}{1} (\bibinfo{year}{2018}), \bibinfo{pages}{49--59}.
\newblock
\urldef\tempurl%
\url{https://doi.org/10.1109/MCC.2018.011791714}
\showDOI{\tempurl}


\bibitem[\protect\citeauthoryear{Richardson}{Richardson}{[n.d.]a}]%
        {RichardsonCQRSPattern}
\bibfield{author}{\bibinfo{person}{C. Richardson}.}
  \bibinfo{year}{[n.d.]}\natexlab{a}.
\newblock \bibinfo{booktitle}{\emph{Command Query Responsibility Segregation
  Pattern}}.
\newblock
\urldef\tempurl%
\url{ttps://microservices.io/patterns/data/cqrs.html}
\showURL{%
Retrieved Feb 05, 2021 from \tempurl}


\bibitem[\protect\citeauthoryear{Richardson}{Richardson}{[n.d.]b}]%
        {RichardsonEventSourcingPattern}
\bibfield{author}{\bibinfo{person}{C. Richardson}.}
  \bibinfo{year}{[n.d.]}\natexlab{b}.
\newblock \bibinfo{booktitle}{\emph{Event Sourcing Pattern}}.
\newblock
\urldef\tempurl%
\url{https://microservices.io/patterns/data/event-sourcing.html}
\showURL{%
Retrieved Feb 05, 2021 from \tempurl}


\bibitem[\protect\citeauthoryear{Richardson}{Richardson}{[n.d.]c}]%
        {richardson_microservices_2018}
\bibfield{author}{\bibinfo{person}{C. Richardson}.}
  \bibinfo{year}{[n.d.]}\natexlab{c}.
\newblock \bibinfo{booktitle}{\emph{Microservices Patterns: With examples in
  Java}}.
\newblock \bibinfo{publisher}{Manning Publications}.
\newblock
\showISBNx{978-1-61729-454-9}
\showLCCN{2018289404}
\urldef\tempurl%
\url{https://books.google.de/books?id=UeK1swEACAAJ}
\showURL{%
\tempurl}


\bibitem[\protect\citeauthoryear{Richardson}{Richardson}{[n.d.]d}]%
        {RichardsonSagaPattern}
\bibfield{author}{\bibinfo{person}{C. Richardson}.}
  \bibinfo{year}{[n.d.]}\natexlab{d}.
\newblock \bibinfo{booktitle}{\emph{Saga Pattern}}.
\newblock
\urldef\tempurl%
\url{https://microservices.io/patterns/data/saga.html}
\showURL{%
Retrieved Feb 05, 2021 from \tempurl}


\bibitem[\protect\citeauthoryear{Richardson}{Richardson}{[n.d.]e}]%
        {RichardsonOutboxPattern}
\bibfield{author}{\bibinfo{person}{C. Richardson}.}
  \bibinfo{year}{[n.d.]}\natexlab{e}.
\newblock \bibinfo{booktitle}{\emph{Transactional Outbox Pattern}}.
\newblock
\urldef\tempurl%
\url{https://microservices.io/patterns/data/transactional-outbox.html}
\showURL{%
Retrieved Feb 05, 2021 from \tempurl}


\bibitem[\protect\citeauthoryear{Scavuzzo, Nitto, and Ardagna}{Scavuzzo
  et~al\mbox{.}}{2018}]%
        {DBLP:journals/ese/ScavuzzoNA18}
\bibfield{author}{\bibinfo{person}{Marco Scavuzzo},
  \bibinfo{person}{Elisabetta~Di Nitto}, {and} \bibinfo{person}{Danilo
  Ardagna}.} \bibinfo{year}{2018}\natexlab{}.
\newblock \showarticletitle{Experiences and challenges in building a data
  intensive system for data migration}.
\newblock \bibinfo{journal}{\emph{Empir. Softw. Eng.}} \bibinfo{volume}{23},
  \bibinfo{number}{1} (\bibinfo{year}{2018}), \bibinfo{pages}{52--86}.
\newblock
\urldef\tempurl%
\url{https://doi.org/10.1007/s10664-017-9503-7}
\showDOI{\tempurl}


\bibitem[\protect\citeauthoryear{Schneider}{Schneider}{1990}]%
        {DBLP:journals/csur/Schneider90}
\bibfield{author}{\bibinfo{person}{Fred~B. Schneider}.}
  \bibinfo{year}{1990}\natexlab{}.
\newblock \showarticletitle{Implementing Fault-Tolerant Services Using the
  State Machine Approach: {A} Tutorial}.
\newblock \bibinfo{journal}{\emph{{ACM} Comput. Surv.}} \bibinfo{volume}{22},
  \bibinfo{number}{4} (\bibinfo{year}{1990}), \bibinfo{pages}{299--319}.
\newblock
\urldef\tempurl%
\url{https://doi.org/10.1145/98163.98167}
\showDOI{\tempurl}


\bibitem[\protect\citeauthoryear{Shapiro, Pregui{\c{c}}a, Baquero, and
  Zawirski}{Shapiro et~al\mbox{.}}{2011}]%
        {DBLP:conf/sss/ShapiroPBZ11}
\bibfield{author}{\bibinfo{person}{Marc Shapiro}, \bibinfo{person}{Nuno~M.
  Pregui{\c{c}}a}, \bibinfo{person}{Carlos Baquero}, {and}
  \bibinfo{person}{Marek Zawirski}.} \bibinfo{year}{2011}\natexlab{}.
\newblock \showarticletitle{Conflict-Free Replicated Data Types}. In
  \bibinfo{booktitle}{\emph{Stabilization, Safety, and Security of Distributed
  Systems - 13th International Symposium, {SSS} 2011, Grenoble, France, October
  10-12, 2011. Proceedings}} \emph{(\bibinfo{series}{Lecture Notes in Computer
  Science})}, \bibfield{editor}{\bibinfo{person}{Xavier D{\'{e}}fago},
  \bibinfo{person}{Franck Petit}, {and} \bibinfo{person}{Vincent Villain}}
  (Eds.), Vol.~\bibinfo{volume}{6976}. \bibinfo{publisher}{Springer},
  \bibinfo{pages}{386--400}.
\newblock
\showISBNx{978-3-642-24549-7}
\urldef\tempurl%
\url{https://doi.org/10.1007/978-3-642-24550-3\_29}
\showDOI{\tempurl}


\bibitem[\protect\citeauthoryear{Staron}{Staron}{2020}]%
        {DBLP:books/sp/Staron20}
\bibfield{author}{\bibinfo{person}{Miroslaw Staron}.}
  \bibinfo{year}{2020}\natexlab{}.
\newblock \bibinfo{booktitle}{\emph{Action Research in Software Engineering -
  Theory and Applications}}.
\newblock \bibinfo{publisher}{Springer}.
\newblock
\showISBNx{978-3-030-32609-8}
\urldef\tempurl%
\url{https://doi.org/10.1007/978-3-030-32610-4}
\showDOI{\tempurl}


\bibitem[\protect\citeauthoryear{Terry}{Terry}{2008}]%
        {DBLP:series/synthesis/2008Terry}
\bibfield{author}{\bibinfo{person}{Douglas~B. Terry}.}
  \bibinfo{year}{2008}\natexlab{}.
\newblock \bibinfo{booktitle}{\emph{Replicated Data Management for Mobile
  Computing}}.
\newblock \bibinfo{publisher}{Morgan {\&} Claypool Publishers}.
\newblock
\urldef\tempurl%
\url{https://doi.org/10.2200/S00132ED1V01Y200807MPC005}
\showDOI{\tempurl}


\bibitem[\protect\citeauthoryear{Vernon}{Vernon}{2013}]%
        {vernon2013implementingDDD}
\bibfield{author}{\bibinfo{person}{Vaughn Vernon}.}
  \bibinfo{year}{2013}\natexlab{}.
\newblock \bibinfo{booktitle}{\emph{Implementing domain-driven design}}.
\newblock \bibinfo{publisher}{Addison-Wesley}.
\newblock


\bibitem[\protect\citeauthoryear{Vogels}{Vogels}{2009}]%
        {DBLP:journals/cacm/Vogels09}
\bibfield{author}{\bibinfo{person}{Werner Vogels}.}
  \bibinfo{year}{2009}\natexlab{}.
\newblock \showarticletitle{Eventually consistent}.
\newblock \bibinfo{journal}{\emph{Commun. {ACM}}} \bibinfo{volume}{52},
  \bibinfo{number}{1} (\bibinfo{year}{2009}), \bibinfo{pages}{40--44}.
\newblock
\urldef\tempurl%
\url{https://doi.org/10.1145/1435417.1435432}
\showDOI{\tempurl}


\bibitem[\protect\citeauthoryear{Weikum and Schek}{Weikum and Schek}{1992}]%
        {DBLP:books/mk/elmagarmid92/WeikumS92}
\bibfield{author}{\bibinfo{person}{Gerhard Weikum} {and}
  \bibinfo{person}{Hans{-}J{\"{o}}rg Schek}.} \bibinfo{year}{1992}\natexlab{}.
\newblock \showarticletitle{Concepts and Applications of Multilevel
  Transactions and Open Nested Transactions}.
\newblock See \citeN{DBLP:books/mk/Elmagarmid92}, \bibinfo{pages}{515--553}.
\newblock
\showISBNx{1-55860-214-3}


\bibitem[\protect\citeauthoryear{Wolff}{Wolff}{2016}]%
        {wolff2016microservices}
\bibfield{author}{\bibinfo{person}{Eberhard Wolff}.}
  \bibinfo{year}{2016}\natexlab{}.
\newblock \bibinfo{booktitle}{\emph{Microservices: flexible software
  architecture}}.
\newblock \bibinfo{publisher}{Addison-Wesley Professional}.
\newblock


\end{thebibliography}

\appendix

\end{document}